\documentclass[%
reprint,
showkeys,
superscriptaddress,
preprintnumbers,
nofootinbib,
amsmath,amssymb,
aps,
prd,
floatfix,
]{revtex4-2}
\usepackage{afterpage}
\usepackage{subfigure}
\usepackage{url}
\usepackage{rotating} 
\usepackage{booktabs}
\usepackage{graphicx}
\usepackage{dcolumn}
\usepackage{bm}
\usepackage{float}
\usepackage{amsmath}
\usepackage{hyperref}

\usepackage{xcolor}

\begin{document}
\title{Physics informed operator learning of parameter dependent spectra}

\author{Haohao Gu}
\thanks{These authors contributed equally to the work. \\ \href{mailto:hanlin@stu.pku.edu.cn}{hanlin@stu.pku.edu.cn}}
 \affiliation{Baidu Inc., Beijing 100085, 
 China}
 
\author{Sensen He}
\thanks{These authors contributed equally to the work. \\ \href{mailto:hanlin@stu.pku.edu.cn}{hanlin@stu.pku.edu.cn}}
\affiliation{Baidu Inc., Beijing 100085, 
China}
 
\author{Hanlin Song}
\thanks{These authors contributed equally to the work. \\ \href{mailto:hanlin@stu.pku.edu.cn}{hanlin@stu.pku.edu.cn}}
\affiliation{School of Physics, Peking University, Beijing 100871, China}

\author{Bo Liang}
\email{liangbo22@mails.ucas.ac.cn}
 \affiliation{Center for Gravitational Wave Experiment, National Microgravity Laboratory, Institute of Mechanics, Chinese Academy of Sciences, Beijing 100190, China}
 \affiliation{Taiji Laboratory for Gravitational Wave Universe (Beijing/Hangzhou), University of Chinese Academy of Sciences (UCAS), Beijing 100049, China}
 
\author{Zhenwei Lyu}
\email{zwlyu@dlut.edu.cn}
\affiliation{Leicester International Institute, Dalian University of Technology, Panjin 124221, China}

\author{Xiaoguang Hu}
\affiliation{Baidu Inc., Beijing 100085, China}

\author{Minghui Du}
\affiliation{Center for Gravitational Wave Experiment, National Microgravity Laboratory, Institute of Mechanics, Chinese Academy of Sciences, Beijing 100190, China}

\author{Peng Xu}
 \email{xupeng@imech.ac.cn}
\affiliation{Center for Gravitational Wave Experiment, National Microgravity Laboratory, Institute of Mechanics, Chinese Academy of Sciences, Beijing 100190, China}
 \affiliation{Key Laboratory of Gravitational Wave Precision Measurement of Zhejiang Province, Hangzhou Institute for Advanced Study, UCAS, Hangzhou 310024, China}
 \affiliation{Taiji Laboratory for Gravitational Wave Universe (Beijing/Hangzhou), University of Chinese Academy of Sciences (UCAS), Beijing 100049, China}
 \affiliation{Lanzhou Center of Theoretical Physics, Lanzhou University, Lanzhou 730000, China}

\author{Bo-Qiang Ma}
\email{mabq@pku.edu.cn}
\affiliation{School of Physics, Zhengzhou University, Zhengzhou 450001, China}
\affiliation{School of Physics, Peking University, Beijing 100871, China}
		
\begin{abstract}
Spectral problems governed by differential operators underpin a wide range of physical systems, yet remain computationally challenging because their spectra depend sensitively on continuous parameters and often demand repeated evaluations across parameter space. 
Here we present \texttt{DeepOPiraKAN}, an open source physics informed neural network architecture for spectral analysis.
By combining operator learning with enhanced optimization stability, it captures the underlying parameter-to-spectrum mapping in a single model, avoiding repeated spectral solutions at isolated points in parameter space. As a representative and stringent benchmark, we apply this framework to the computation of quasinormal modes of Kerr black holes.
A single trained network accurately resolves modes with $(\ell,m)\in \{(2,0),(2,1)\}$ and overtones up to $n=7$ across the full spin range, achieving relative errors of $\mathcal{O}(10^{-6})$ for the fundamental mode and gradually increasing to $\mathcal{O}(10^{-4})$ for higher overtones, benchmarked against the Leaver's method. This level of accuracy is already significant for black hole spectroscopy and practical ringdown modelling for current and future observatories. More broadly, these results highlight the potential of \texttt{DeepOPiraKAN} as a general and scalable framework for parameter dependent spectral problems across complex physical systems.
\end{abstract}

\maketitle

\section{Introduction}
Spectral problems occupy a central place in modern mathematical physics, providing the mathematical language for quantization, stability and resonance. Their importance extends across a wide range of problems in physics and engineering, including quantum dynamics, astrophysical oscillations, continuum instabilities and wave propagation in complex structures~\cite{hernandez2002slepc}. However, accurate spectral computation remains a longstanding challenge. In many realistic systems, spectra depend sensitively on continuous physical parameters, making parameter dependent spectral analysis a fundamental yet computationally demanding task. This difficulty is especially pronounced in problems involving strong interactions, complex boundary conditions, or highly constrained geometries, where repeated high-precision eigenvalue solves become both costly and numerically delicate. New approaches to parameter dependent spectral computation could therefore have broad impact across both science and engineering, with potential downstream influence on the spectral algorithms embedded in widely used simulation platforms such as ANSYS and COMSOL~\cite{roman2016slepc}.

Astrophysical compact objects pose a particularly demanding class of parameter dependent spectral problems at the intersection of fundamental physics and scientific computing. In these systems, characteristic oscillation modes encode essential information about strong field gravity. For neutron stars, normal mode oscillations probe the equation of state and internal composition~\cite{Kokkotas:1999bd}, while for black holes, quasinormal modes (QNMs) describe the characteristic response of spacetime to external perturbations~\cite{Konoplya:2011qq,Berti:2009kk}. Governed by non-Hermitian operators with dissipative boundary conditions, QNMs possess complex spectra that are inherently sensitive to the background geometry and, in some cases, to environmental perturbations~\cite{Barausse:2014tra}. This sensitivity makes them a uniquely powerful probe for testing the Kerr no-hair theorem, constraining possible departures from general relativity, investigating the nature of black holes and other ultracompact celestial objects, and probing the properties and spatial distribution of dark matter around massive black holes~\cite{Franchini:2023eda, Berti:2025hly}. These scientific opportunities have made the detection and interpretation of QNMs a major objective of gravitational wave astronomy. They are targeted by current LIGO–Virgo–KAGRA analyses~\cite{LIGOScientific:2016aoc} and form a core part of the science case for future space based observatories such as LISA~\cite{LISA:2017pwj}, Taiji~\cite{Hu:2017mde,Du:2025xdq}, and TianQin~\cite{TianQin:2015yph}, where high-signal-to-noise ringdown signals from massive black hole mergers are expected to enable much sharper spectroscopic tests. In this context, multimode black hole spectroscopy requires going beyond the dominant $(2,2)$ family to include subdominant modes such as $(2,1)$ and $(2,0)$, which carry complementary information about the post-merger black hole spacetime. This, in turn, places stringent demands on the rapid and accurate computation of QNM spectra across parameter space for large scale waveform construction.

Physics informed neural networks (PINNs) have recently emerged as a promising framework for scientific machine learning and differential equations by embedding governing physical laws directly into neural network training~\cite{Raissi:2017zsi,anitescu_2023}. They have already been applied to astrophysical spectral problems, including neutron star oscillations and black hole QNMs in Schwarzschild, Kerr, and extended gravity settings~\cite{Tseneklidou:2025stn,Cornell:2022enn,Patel:2024wzo,Luna:2022rql,Cornell:2024azz,Pombo:2025urp,Luna:2024spo}. However, most existing PINN based spectral solvers remain pointwise rather than operator level, so spectra at different parameter values must be recomputed separately. This limits scalability and is particularly problematic for higher overtones, whose solutions are more sensitive to boundary conditions and numerical precision. As a result, existing methods remain focused primarily on individual modes or narrow parameter regimes, leaving fast, robust, and accurate parameter dependent spectral computation largely unresolved.

To address these limitations, we develop an open source physics informed neural network architecture, termed the Physics Informed Deep Operator Network with Residual Adaptive Kolmogorov–Arnold Network Architecture (\texttt{DeepOPiraKAN}\footnote{The \texttt{DeepOPiraKAN} code developed in this work have been made publicly available through the open source \texttt{PaddleScience} GitHub repository \cite{ppsci}}), for parameter dependent spectral problems. The central idea of \texttt{DeepOPiraKAN} is to elevate spectral learning from a pointwise, instance based paradigm to an operator level formulation that directly maps physical parameters to spectral solutions within a single unified model. It combines the operator learning capability of \texttt{DeepONet}~\cite{Lu_2021,jin_2022} with \texttt{PirateKAN}, a new residual adaptive architecture introduced here to improve optimization stability in spectral regimes with strong parameter sensitivity and increasing structural complexity, such as the higher overtones of Kerr black hole QNMs. By embedding parameters such as black hole spin and overtone number into the branch network, the model captures global spectral structure across continuous parameter space. As a representative and stringent benchmark of this general framework, we apply \texttt{DeepOPiraKAN} to Kerr black hole QNMs, where a single trained network accurately resolves the physically relevant subdominant modes with $lm \in \{20,21\}$ and overtones up to $n=7$ over the full spin range, achieving accuracies several orders of magnitude better than existing PINN based methods. \texttt{DeepOPiraKAN} therefore establishes a scalable and robust framework for parameter dependent spectral learning, with potential applications extending beyond black hole QNMs to a broad class of spectral problems in physics and engineering.

\section{DeepOPiraKAN Architecture}

\begin{figure*}[htbp]
    \centering
    \includegraphics[width=0.9\textwidth]{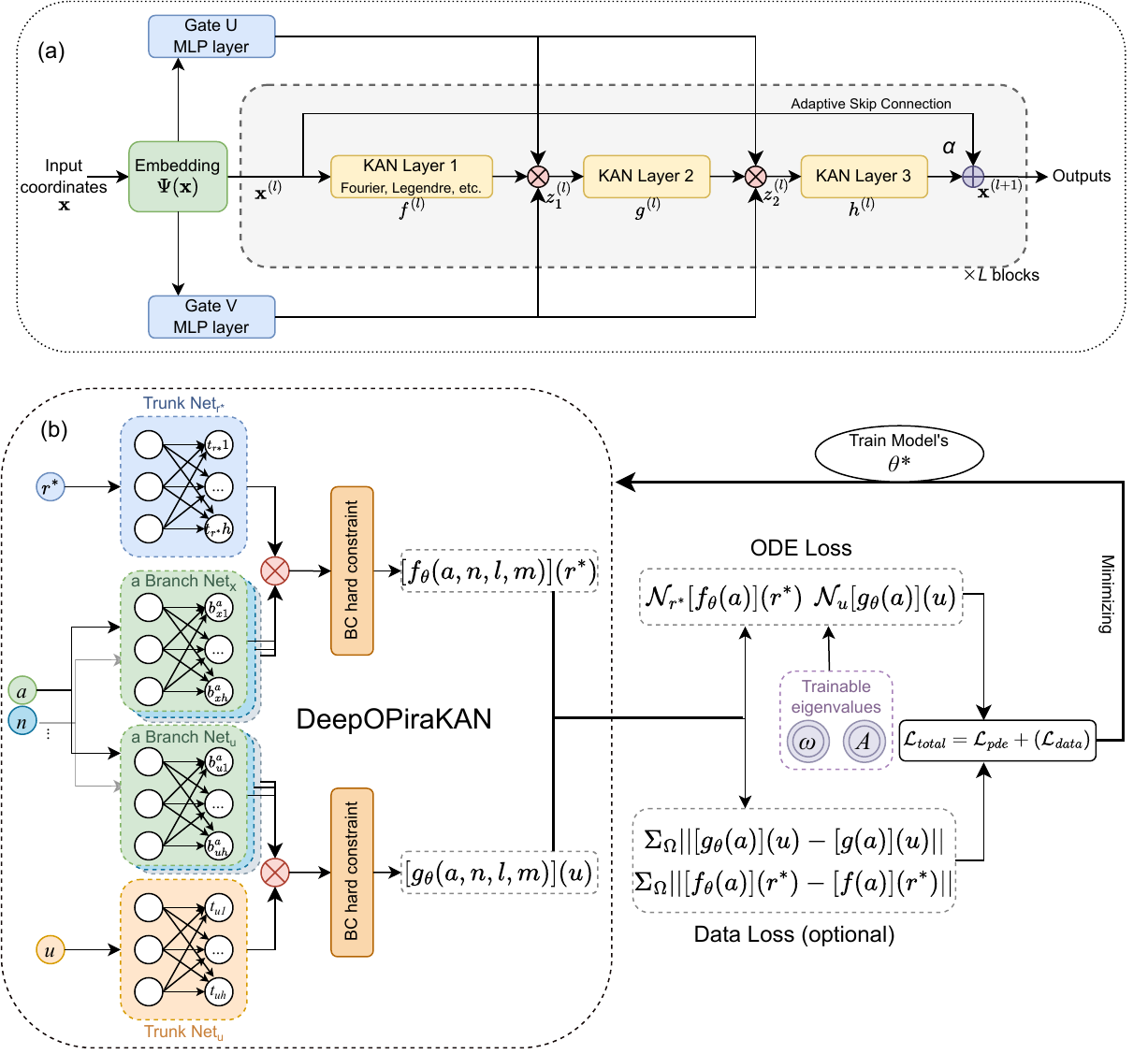}
    \caption{The Physics Informed Deep Operator Network with Residual Adaptive KAN (\texttt{DeepOPiraKAN}) Architecture.
        (a) Physics Informed Residual Adaptive Kolmogorov Arnold Network (\texttt{PiraKAN}): 
        In the current model, input coordinates are first projected into a high dimensional feature space with random Fourier embedding, then passes through \textit{N} blocks of residual adaptive KAN. 
        Each block consists of three \texttt{KAN} layers, gated by two linear dense layers, and adaptively short connected at the end (controlled by a trainable parameter $\alpha$). 
        (b) The multi-branch \texttt{DeepONet} of the current work to solve QNM equations. The radial and angular coordinates serve as input coordinates of two separate trunk nets, which output the base solution. To achieve parameterized solving of QNM, the ODE parameters pass through multi-branch nets, condition the trunk net outputs, and give rise to the final solution. Boundary conditions are imposed as hard constraints on the network outputs. 
    }
    \label{architecture}
\end{figure*}

The application of PINNs to complex nonlinear ordinary differential equation
(ODEs) and partial differential equation (PDEs) is often hindered by severe training pathologies, including spectral bias~\cite{Wang_2021}, unbalanced objectives~\cite{wang_soap_2025}, trivial solutions~\cite{pmlr-v202-daw23a}, and nonphysical local minima~\cite{GOPAKUMAR2023100464, rathore2024challengestrainingpinnsloss}. These difficulties arise primarily from optimization challenges induced by high order differentiation in the loss function, rather than from insufficient approximation capacity.
To overcome this, we propose the Physics Informed Residual Adaptive Kolmogorov Arnold Network (\texttt{PiraKAN}) architecture, which combines the strong nonlinear approximation capacity of the Kolmogorov Arnold Network (\texttt{KAN})~\cite{liu2025kankolmogorovarnoldnetworks, Wang_2025} with the residual adaptive short connection mechanism introduced by Wang {\it et al.}~\cite{wang2024piratenetsphysicsinformeddeeplearning}. This architecture achieves high accuracy and stable convergence when solving 
spectra problems.
Furthermore, we extend the framework by integrating \texttt{PiraKAN} into the \texttt{DeepONet} architecture, replacing both the branch and trunk networks. The resulting \texttt{DeepOPiraKAN} model enables efficient learning of parameter dependent spectral problems at the operator level. Once trained, it can infer solution functions and eigenvalues at given PDE parameters with minimal finetuning. In the following, we describe the proposed architecture in detail.

{\textbf{PiraKAN}.---} The subfigure (a) of Fig.~\ref{architecture} shows the modules of the forward pass of a \texttt{PiraKAN} block. First, the input coordinates $\mathbf{x}\in \mathbb{R}^d$ are projected to a high dimensional ($n$-dim) feature space by the random Fourier embedding function \cite{tancik2020fourierfeaturesletnetworks}, which has been proved to be useful in alleviating spectral bias and better approximation of high frequency solutions: 
\begin{equation}
    \Psi(\mathbf{x}) =
\begin{bmatrix}
\cos(\mathbf{B}\mathbf{x}) \\[3pt]
\sin(\mathbf{B}\mathbf{x})
\end{bmatrix}
\in \mathbb{R}^{n},
\end{equation}
where $\mathbf{B}\in\mathbb{R}^{n/2\times d}$ has entities i.i.d. sampled from Gaussian distribution $\mathcal{N}(0, s ^ 2)$ with the standard deviation $s$ set as a hyperparameter.

After the projection, the embedded coordinates $\Psi(\mathbf{x})$ are fed into two dense layers $\mathbf{U}$, $\mathbf{V}$, both follow standard feedforward linear layer formulation and serve as gates in the blocks:
\begin{equation}
    \mathbf{U} =
\sigma(
\mathbf{W_u}\Psi(\mathbf{x}))
+ \mathbf{b_u},
\mathbf{V} =
\sigma(
\mathbf{W_v}\Psi(\mathbf{x}))
+ \mathbf{b_v}
,
\end{equation}
where $\sigma$ is a pointwise activation function. Such design has been widely employed to enhance the trainability and convergence of PINNs \cite{wang2020understandingmitigatinggradientpathologies, anagnostopoulos2023residualbasedattentionconnectioninformation, wang2024piratenetsphysicsinformeddeeplearning}.

Let $\mathbf{x^{(l)}}$ be the input of the $l$-th block of \texttt{PiraKAN} ($1 \leq l \leq L$), the feed forward process can be formulated as follows:
\begin{equation}
f^{(l)} = 
\sum_{q=1}^{2n+1} 
\Phi_{1q}^{(l)} 
\left( 
\sum_{p=1}^{n} 
\phi_{1q,p}^{(l)} 
(x_p) 
\right),
\end{equation}

\begin{equation}
    \mathbf{z}_1^{(l)} 
    = f^{(l)} 
    \odot \mathbf{U} 
    + (1 - f^{(l)}) 
    \odot \mathbf{V},
\end{equation}

\begin{equation}
    g^{(l)} = 
    \sum_{q=1}^{2n+1} 
    \Phi_{2q}^{(l)} 
    \left( 
    \sum_{p=1}^{n} 
    \phi_{2q,p}^{(l)} 
    (z_{1p}^{(l)}) 
    \right),
\end{equation}

\begin{equation}
    \mathbf{z}_2^{(l)} 
    = g^{(l)} 
    \odot \mathbf{U} 
    + (1 - g^{(l)}) 
    \odot \mathbf{V},
\end{equation}

\begin{equation}
    h^{(l)} = 
    \sum_{q=1}^{2n+1} 
    \Phi_{3q}^{(l)} 
    \left( 
        \sum_{p=1}^{n} 
        \phi_{3q,p}^{(l)} 
        (z_{2p}^{(l)})
    \right),
\end{equation}

\begin{equation}
    \label{transfer_function}
    \mathbf{x}^{(l+1)} = 
    \alpha^{(l)} h^{(l)} 
    + (1 - \alpha^{(l)}) 
    \mathbf{x}^{(l)},
\end{equation}
where $\odot$ denotes the Hadamard (elementwise) product, and $\phi_{iq,p}^{(l)}$ is a single variable nonlinear function of the $p$-th entry of block input $\mathbf{x^{(l)}}$, which serves as the `edge' function in \texttt{KAN} architecture \cite{liu2025kankolmogorovarnoldnetworks}, the lower script $i = 1, 2, 3$ denotes the $i$-th \texttt{KAN} layer of the $l$-th block. In current work, we employ Fourier series $\phi(x) = a_0 + \sum_{m=1}^{M}(a_m \cos(m \omega x) + b_m \sin(m \omega x) )$ as the edge function \cite{zhang2025kolmogorovarnoldfouriernetworks, farea2025learnableactivationfunctionsphysicsinformed}, where $M$, the cutoff frequency mode, is a user defined hyperparameter. 

All the weights are initialized by the Glorot Normal scheme, while the biases are initialized to 0. The $\alpha^{(l)}\in \mathbb{R}$ in Eq.~(\ref{transfer_function}) is a trainable parameter which controls the nonlinear output ratio of each block. During initialization, $\alpha^{(l)}$ are set to 0, which means the network reduces to linear mapping. With training proceeds, the $\alpha^{(l)}$ gradually increases in a layerwised manner (the shallow blocks are activated before the deeper blocks). This design introduced in Ref.~\cite{wang2024piratenetsphysicsinformeddeeplearning} has been proved useful to circumvent the initialization pathologies of PINNs at the beginning stage of training, largely prevent the NNs from convergence to trivial solutions.
Additionally, the nonlinear approximation capacity endowed by Fourier \texttt{KAN} \cite{zhang2025kolmogorovarnoldfouriernetworks, farea2025learnableactivationfunctionsphysicsinformed} greatly reduces the depth and width to achieve similar expressiveness compared with multilayer perceptrons (MLPs). Consequently, the total number of trainable parameters required to simulate the solution functions of various PDEs are largely reduced. Such a feature simplifies the optimization task for PINNs training, and leads to faster and more stable convergence in PDE-solving tasks.

{\textbf{DeepONet}.---}To enable fast inference of parameterized PDE solutions at the operator level, we adopt the \texttt{DeepONet} framework \cite{Lu_2021,jin_2022}, in which an operator 
$\mathcal{G}: u(\cdot,\boldsymbol{\mu}) \mapsto y(\cdot,\boldsymbol{\mu})$
is learned either from input-output function pairs 
$\{(u_i(\mathbf{x}), y_i(\mathbf{x}))\}_{i=1}^{N}$ or from PDE residuals \cite{doi:10.1126/sciadv.abi8605}.
The \texttt{DeepONet} represents the target operator as an inner product between the latent feature of one or multiple input function(s) $u(\cdot)$ (encoded by the \emph{branch} network) and that of a coordinate point $\mathbf{x}$ (encoded by the \emph{trunk} network)\cite{jin_2022}:
\begin{equation}
    {\mathcal{G}}(\mathbf{x}|u_{1...n})
    = \mathcal{S} \left(
        \underbrace{{B}_1(u_1)}_{\text{branch}_1}
        \odot \cdots \odot
        \underbrace{{B}_n(u_n)}_{\text{branch}_n}
        \odot
        \underbrace{T_k(\mathbf{x})}_{\text{trunk}}
    \right) + b,
    \label{eq:deeponet}
\end{equation}
where $\mathcal{S}$ represents the summation over all components of a vector, and $b \in \mathbb{R}$ is a trainable bias. The term ${B}_1(u_1)$ denotes the encoded feature of the $i$-th input function or PDE parameter through the $i$-th branch network, while $T_k(\mathbf{x})$ denotes the encoded coordinate via the trunk network.
Once trained, the \texttt{DeepONet} can evaluate the solution $y(\mathbf{x},\boldsymbol{\mu})$ for arbitrary input function $u(\cdot)$ or PDE parameters $\boldsymbol{\mu}$ in a single forward pass, enabling rapid inference.

{\textbf{DeepOPiraKAN}.---}In the proposed \texttt{DeepOPiraKAN} architecture, both the branch and trunk networks are implemented with the \texttt{PiraKAN} blocks introduced in the previous section. Specifically, each branch network $\mathcal{B}_{\theta_b}$ maps one PDE parameter to a latent feature vector:
\begin{equation}
 \mathbf{B}^{\mu}(\boldsymbol{\mu}) = \mathcal{B}^{\mu}_{\theta_b}(\boldsymbol{\mu}) \in \mathbb{R}^{r},
\end{equation}
where $\boldsymbol{\mu}$ denotes the PDE parameters (\emph{e.g.} spin and mode numbers when solving the Teukolsky equation).
while the trunk network $\mathcal{T}_{\theta_t}$ encodes the spatial temporal coordinates:
\begin{equation}
    \mathbf{T}(\mathbf{x}) = \mathcal{T}_{\theta_t}(\mathbf{x}) \in \mathbb{R}^{r}.
\end{equation}
The predicted solution is then obtained through Eq.~(12).
Both $\mathcal{B}_{\theta_b}$ and $\mathcal{T}_{\theta_t}$ are implemented by multiple \texttt{PiraKAN} blocks. This design enables strong nonlinear representation while maintaining stable optimization properties for physics informed learning. In addition, the eigenvalues of the governing equations are incorporated into the trainable parameters, allowing both the forward problem (PDE solving) and the inverse problem (parameter identification) to be addressed simultaneously. 

During training, the boundary conditions can be imposed either as \textit{hard constraints} directly on the network outputs, or as \textit{soft constraints} formulated as the terms of loss functions. The overall loss function combines physics based constraints and supervised data constraints (if input-output function data pairs are available):
\begin{equation}
\mathcal{L} = \mathcal{L}_{\mathrm{data}}
+ \lambda_{\mathrm{pde}} \mathcal{L}_{\mathrm{PDE}}
+ \lambda_{\mathrm{bc}} \mathcal{L}_{\mathrm{BC}},
\end{equation}
where $\mathcal{L}_{\mathrm{PDE}}$ enforces the governing equation residuals, and $\mathcal{L}_{\mathrm{BC}}$ ensures boundary or initial conditions are satisfied.

Using \texttt{PiraKAN} as the building block of \texttt{DeepONet} offers three main advantages. The combination of Fourier feature embedding and KAN based edge functions improves the representation of multiscale PDE solutions and mitigates spectral bias. The residual adaptive short connections, governed by trainable coefficients $\alpha^{(l)}$, enhance optimization stability by reducing the initialization pathologies often observed in PINNs and related operator learning architectures. In addition, the Kolmogorov–Arnold functional decomposition increases expressive efficiency, enabling comparable approximation accuracy with reduced depth and width, which in turn lowers parameter counts and supports faster, more stable convergence.
Once trained, the \texttt{DeepOPiraKAN} can quickly predict the solution field $y(\mathbf{x},\boldsymbol{\mu})$ and associated eigenvalues for new PDE parameters $\boldsymbol{\mu}$ by finetuning of the eigenvalues instead of resolving the governing equations.

\section{The Quasinormal modes of Kerr Black hole}
\label{qnms_Teukolsky_Eqs}
The QNMs denotes the characteristic oscillations of perturbed black holes, determined solely by their masses, spins, and charges~\cite{Kokkotas:1999bd}. The QNMs represent the intrinsic ``fingerprints'' of black holes, encoding the fundamental properties of the general relativity. The first detection of ringdown stage by LIGO–Virgo–KAGRA Collaboration has confirmed the existence of QNMs~\cite{LIGOScientific:2016aoc}, with the dominate mode observed in the gravitational wave of ringdown signal of GW150914~\cite{LIGOScientific:2016lio}.  Moreover, the high order ringdown modes are believed to have been observed in GW231123~\cite{LIGOScientific:2025rsn} and GW250114~\cite{LIGOScientific:2025obp}. Future detectors like LISA~\cite{LISA:2017pwj}, Taiji~\cite{  Hu:2017mde, Du:2025xdq}, TianQin~\cite{TianQin:2015yph}, Cosmic Explorer~\cite{Reitze:2019iox} and Einstein telescope~\cite{Punturo:2010zz} are expected to resolve more multiple subdominant modes, enabling 
black hole spectroscopy to test the Kerr hypothesis and probe GR in the strong field regime~\cite{Berti:2007fi, Bhagwat:2021kwv}. Accurate and efficient computation of QNM spectra is therefore essential for interpreting gravitational wave observations. 

The gravitational perturbations of a black hole with mass $M$ and spin $a$ are governed by the Teukolsky equation \cite{Teukolsky:1972my, Teukolsky:1973ha},

\begin{widetext}
\begin{equation}\begin{gathered}\left[\frac{(r^2+a^2)^2}{\Delta}-a^2\sin^2\theta\right]\frac{\partial^2\psi}{c^2\partial t^2}+\frac{4GMar}{c^3\Delta}\frac{\partial^2\psi}{\partial t\partial\phi}+\left[\frac{a^2}{\Delta}-\frac{1}{\sin^2\theta}\right]\frac{\partial^2\psi}{\partial\phi^2}-\Delta^2\frac{\partial}{\partial r}\left(\frac{1}{\Delta}\frac{\partial\psi}{\partial r}\right)-\frac{1}{\sin\theta}\frac{\partial}{\partial\theta}\left(\sin\theta\frac{\partial\psi}{\partial\theta}\right)\\+4\left[\frac{a(r-GM/c^2)}{\Delta}+i\frac{\cos\theta}{\sin^2\theta}\right]\frac{\partial\psi}{\partial\phi}+4\left[\frac{GM(r^2-a^2)/c^2}{\Delta}-r-\mathrm{i}a\cos\theta\right]\frac{\partial\psi}{c\partial t}+2(2\cot^2\theta+1)\psi=0,\end{gathered}\end{equation}
\end{widetext}

\noindent where $\Delta:=~r^{2}- 2GMr/c^{2}+a^{2}$.  After the standard procedure of separation of variables, the perturbation function $\psi$ can be expressed as,

\begin{equation}
    \psi=\sum_{l=2}^{\infty}\sum_{m=-l}^{l}\mathrm{e}^{im\phi}\int_{-\infty}^{\infty}S_{l m}(\cos\theta~,\omega)R_{l m}(r,\omega)\mathrm{e}^{i\omega t}\frac{d\omega}{2\pi},
\end{equation}
where $(t,r,\theta,\phi)$ denote the spacetime coordinates. 
The perturbation function $\psi$ is complex and can be separated into a radial part $R_{lm}(r,\omega)$ and an angular part $S_{lm}(\cos\theta, \omega)$.

\begin{widetext}
\begin{equation}
\label{radial_part}
\begin{gathered}
\Delta^2\frac d{dr}\left(\frac1\Delta\frac{d R_{l m}}{d r}\right)+\left\{\frac{[(r^2+a^2)(\omega/c)-am]^2-4i(r-GM/c^2)[(r^2+a^2)(\omega/c)-am]}\Delta\right.\\-8ir(\omega/c)+2am(\omega/c)-a^2(\omega/c)^2-A_{l m}\Bigg\}R_{l m}=0,
\end{gathered}
\end{equation}

\begin{align}
\begin{array}{l}{{\frac{d}{d\cos\theta}\left[(1-\cos^{2}\theta){\frac{d~S_{l m}}{d\cos\theta}}\right]}} {{+\left[a^{2}(\omega/c)^{2}\cos^{2}\theta+4a(\omega/c)\cos\theta-2-{\frac{(m-2\cos\theta)^{2}}{1-\cos^{2}\theta}}+A_{l m}\right]S_{l m}}} = 0 ,\end{array} 
\end{align}
\end{widetext}

where $A_{lm}$ is the separation constant, and $\omega$ denotes the QNM frequency, with its real part representing the oscillation frequency and its imaginary part characterizing the damping rate of the perturbation.

The computation of QNMs has a long history, with a variety techniques developed to solve the perturbation equation~\cite{Kokkotas:1999bd}. Among them, the Leaver's continued fraction method \cite{Leaver:1985ax} is widely regarded as a benchmark approach for computing QNMs. In this method, the solution to the Teukolsky equation is expressed as an infinite series expansion, whose coefficients satisfy a three term recurrence relation that can be cast into a continued fraction form. Substituting this series into the original equation transforms the differential eigenvalue problem into an algebraic root finding one, significantly improving both computational efficiency and numerical precision. Other approaches, such as evolving the time dependent wave equation \cite{Vishveshwara:1970zz},  evolving the time independent wave equation \cite{Chandrasekhar:1975zza}, and the WKB approximation \cite{Schutz:1985km, Iyer:1986np}, have also been extensively developed and refined over the past decades. While these traditional methods have achieved remarkable accuracy, they typically require solving the eigenvalue problem separately for each set of black hole parameters, making large scale exploration across spin and mode space computationally expensive.

These considerations indicate that the main challenge in QNM calculations is not the accuracy of individual solvers, but the repeated solution of closely related eigenvalue problems across continuous parameter space. This naturally motivates a learning based reformulation, in which the goal is to capture the functional dependence of the spectrum on black hole parameters within a unified framework, rather than recomputing each configuration independently. Within this framework, to facilitate neural network training, the radial function $R_{lm}(r,\omega)$ and angular function $S_{lm}(\cos\theta,\omega)$ are further simplified through the introduction of transformed variables\footnote{Unlike Ref.~\cite{Luna:2022rql}, we adopt the notation $r^* = r_+/r$ instead of $x = r_+/r$ to maintain notational consistency throughout this paper.} $r^* = r_+/r$ and $u = \cos\theta$, where $r_+ = (1 + \sqrt{1-4a^2})/2$. Under this transformation, the domains are mapped to the regular intervals $[0,1]$ and $[-1,1]$, respectively. Following the approach of Luna {\it et al.}~\cite{Luna:2022rql}, the radial and angular equations can then be reduced to the following simplified forms,
\begin{equation}
\begin{aligned}
\mathcal{L}_{F}[f(r^*)]&=F_{2}f^{\prime\prime}(r^*)+F_{1}f^{\prime}(r^*)+F_{0}f(r^*)=0,\\
\mathcal{L}_{G}[g(u)]&=G_{2}g^{\prime\prime}(u)+G_{1}g^{\prime}(u)+G_{0}g(u)=0,
\end{aligned}
\end{equation}
where the convention $G=c=2M=1$ is assumed for simplicity, and $F_{{i}}(r^*)$ and $G_{{i}}(u)$ denote coefficient functions, which can be found in the Appendix of Luna {\it et al.}~\cite{Luna:2022rql}. The associated boundary conditions are then given by,
\begin{equation}
f\left(1\right)=1,\qquad g\left(-1\right)=1.
\end{equation}

\section{Results}
\label{results}


\begin{table*}[ht]
    \centering
    \small 
    \caption{Architectural configurations of \texttt{DeepOPiraKAN}.}
    \begin{tabular}{l l @{\hskip 0.3in} l l @{\hskip 0.3in} l l}
    \toprule
    \multicolumn{2}{c}{\textbf{Architecture}} & \multicolumn{2}{c}{\textbf{Training}} & \multicolumn{2}{c}{\textbf{Evaluation}} \\
    \cmidrule(r){1-2} \cmidrule(lr){3-4} \cmidrule(l){5-6}
    \textbf{Hyperparam.} & \textbf{Value} & \textbf{Hyperparam.} & \textbf{Value} & \textbf{Hyperparam.} & \textbf{Value} \\
    \midrule
    Hidden dim & 64  & Param batch & 100 & Steps & $(2\sim4)\cdot 10^{3}$ \\
    Blocks ($L$) & 2 & PDE batch & 128 & Init LR & $1\times10^{-3}$ \\
    KAN grid & 1 & Init LR & $1\times10^{-3}$ & Decay steps & $4\times10^{2}$ \\
    Activation & Tanh& Steps & $2\times10^{5}$ & & \\
    Fourier scale & 0.7&  Decay rate & 0.9  & & \\
    Weight fact. & $\mu=1.0, \sigma=0.1$  &Decay steps & $5\times10^{3}$ & & \\
    \bottomrule
    \end{tabular}
    \label{tab:reconfigured_kan}
\end{table*}

\subsection{Computational framework and training strategy}
The configurations of \texttt{DeepOPiraKAN} are summarized in Table 1. Both the trunk and branch networks adopt the same \texttt{PiraKAN} configuration, including the hidden dimension, number of blocks, \texttt{KAN} grid size, activation function, and factorized random initialization. The trunk network takes the spatial coordinates $(r^*, u)$ as input, while the branch network encodes the physical parameters $(a, n)$.


During training, each minibatch is constructed by first sampling 100 parameter pairs $(a \in {[0,0.5)}, n \in \{ 0,1,2...,7 \} )$, and then drawing 128 collocation points of $(r^*, u)$ for each parameter pair.
The training dataset is generated from high precision QNM computed with Leaver's method.
For each overtone $n$, eigenvalues are precomputed on a uniform grid of 100 spin values in $a\in[0,0.5)$. During training, 22 spin values are uniformly subsampled for each overtone, resulting in a total of 176 $(a,n)$ eigenvalue labels used to supervise the model. These eigenvalues are embedded into the 
loss function as parametric inputs, enabling the network to learn a continuous operator mapping from the physical parameters $(a,n)$ to the corresponding solution manifold and spectral values.
The loss function includes the Teukolsky equation residual evaluated at these points, while the boundary conditions are hard imposed to the network output following Ref.~\cite{Luna:2022rql}. Training is performed using the \texttt{SOAP}~\cite{wang_soap_2025} optimizer for $2\times10^{5}$ with an exponential decay schedule. The combination of strong nonlinear representation and adaptive short residual connections facilitates the stable optimization, allowing the network to converge uniformly across different spins and overtones. 

After training, all network parameters are frozen, and only the complex eigenvalues $\omega$ and $A$ are treated as learnable parameters during evaluation.  These two parameters are initialized using empirical analytic formulas that depend on the spin $a$ and the mode numbers $lmn$.  
The initial guess for $\omega$ follows Refs.~\cite{Ferrari:1984zz, Yang:2012he, QuasiNormalModesToolkit},
\begin{equation}
    \omega\approx(l+1/2)\Omega-i\gamma_{L}(n+1/2),
\end{equation}
where the $\Omega$ is the Keplerian frequency of the circular photon orbit and $\gamma_{L}$ is the Lyapunov exponet of the orbit. Both quantities also depend on spin $a$ and the mode numbers $lm$. The initial guess for the angular eigenvalue $A$ is taken from Refs.~\cite{Iyer:1986np, Yang:2012he, QuasiNormalModesToolkit},
\begin{equation}
    A_{lm}\approx (l+1/2)^{2}-\frac{a^{2}\omega^{2}}{2}\bigg[1-\frac{m^{2}}{(l+1/2)^{2}}\bigg].
\end{equation}
For a given $(a, n)$, the eigenvalue is optimized for $2\times10^{3}$ or $4\times10^{3}$ steps to minimize the residual loss, yielding the final prediction. 

The development, traing and evalution of current model are implemented using \texttt{PaddlePaddle 3.0}~\cite{paddle} deep learning framework together with \texttt{PaddleScience 1.4} scientific machine learning toolkit~\cite{ppsci}. Both the pretraining and evaluation are performed on a single NVIDIA A100-40G GPU. The pretraining of \texttt{DeepOPiraKAN} takes approximately 19 hours on single A100. During evaluation, the computational cost is determined by the finetuning steps of eigenvalue, with an average runtime of about 70 seconds per $10^3$ optimization steps.

\subsection{Benchmarking on Kerr quasinormal modes}

To demonstrate the practical utility of our framework, we apply \texttt{DeepOPiraKAN} to the Teukolsky equation governing perturbations of Kerr black holes. We study the QNM spectrum in detail for the fundamental mode and the first seven overtones of the $(\ell,m)=(2,0)$ and $(2,1)$ families across the full Kerr spin range. Beyond the well known dominant $(2,2)$ branch, these families are among the subdominant modes most relevant for multimode black hole spectroscopy, while future space based observations are expected to reach signal-to-noise levels at which incomplete mode modeling may become a significant source of systematic error~\cite{Franchini:2023eda,Baibhav:2018rfk,Berti:2025hly}. Resolving these families and their overtones across the full Kerr spin range is therefore directly relevant to the accurate modeling required for future ringdown inference and spectroscopy. More importantly, the results show that \texttt{DeepOPiraKAN} can learn the structured dependence of eigenvalues on continuous physical parameters, establishing a general and reusable framework for parameter dependent spectral computation beyond the specific Kerr setting studied here.

\begin{figure*}[htbp]
\centering
    \begin{minipage}{0.48\textwidth}
        \centering
        \includegraphics[width=\textwidth]{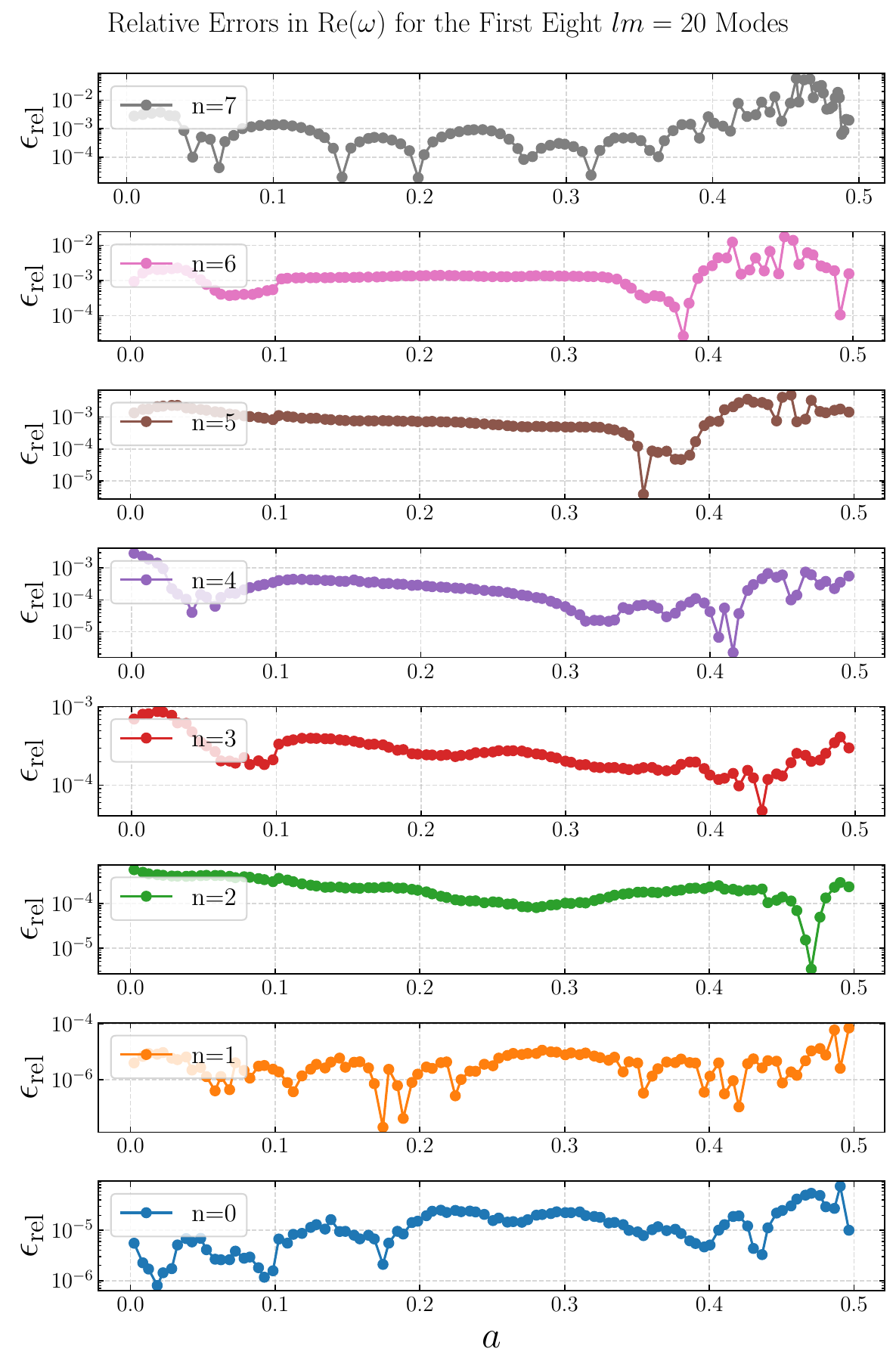}
    \end{minipage}
    \hspace{0.01\textwidth}
    \begin{minipage}{0.48\textwidth}
        \centering
        \includegraphics[width=\textwidth]{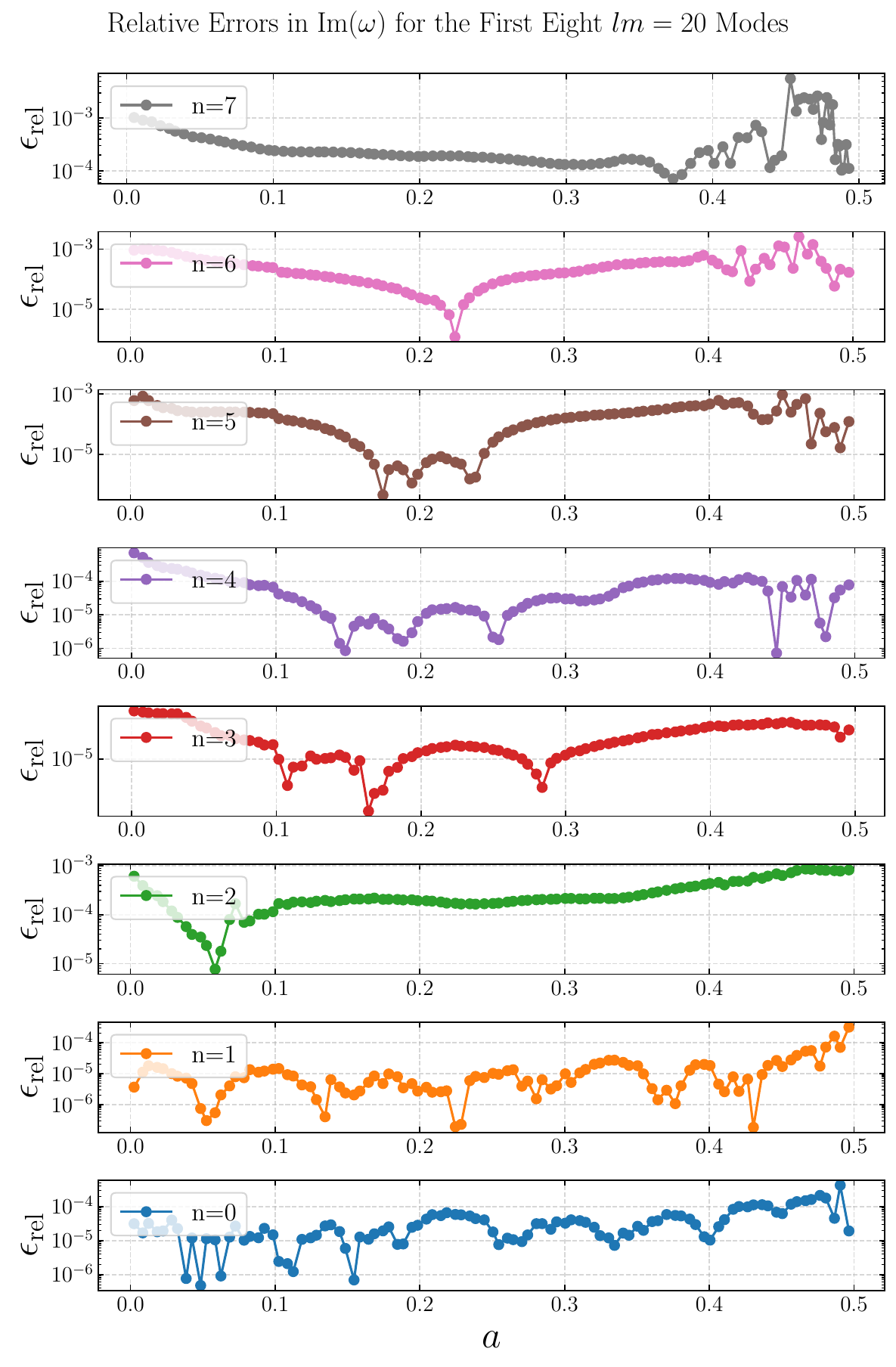}
    \end{minipage}
    \caption{Relative accuracy of the first eight QNMs for the $lm=20$ case across the full spin range obtained with \texttt{DeepOPiraKAN}. The left panel shows the relative error in the real part of the QNM frequencies, while the right panel shows the corresponding error in the imaginary part. Relative errors are evaluated with respect to benchmark values computed using the Leaver's method.}
    \label{lm20} 	
\end{figure*}

\begin{figure*}[htbp]
\centering
    \begin{minipage}{0.48\textwidth}
        \centering
        \includegraphics[width=\textwidth]{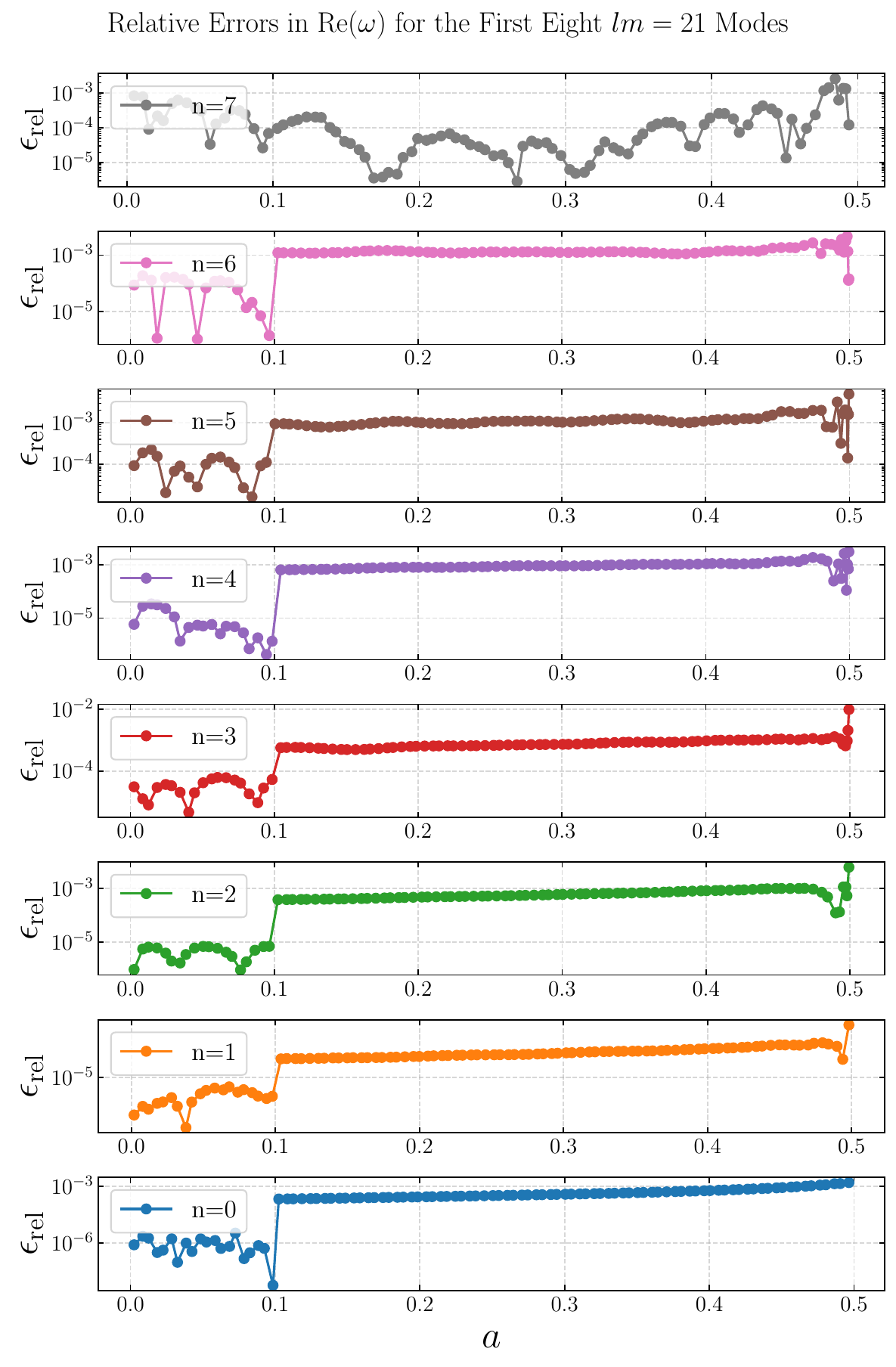}
    \end{minipage}
    \hspace{0.01\textwidth}
    \begin{minipage}{0.48\textwidth}
        \centering
        \includegraphics[width=\textwidth]{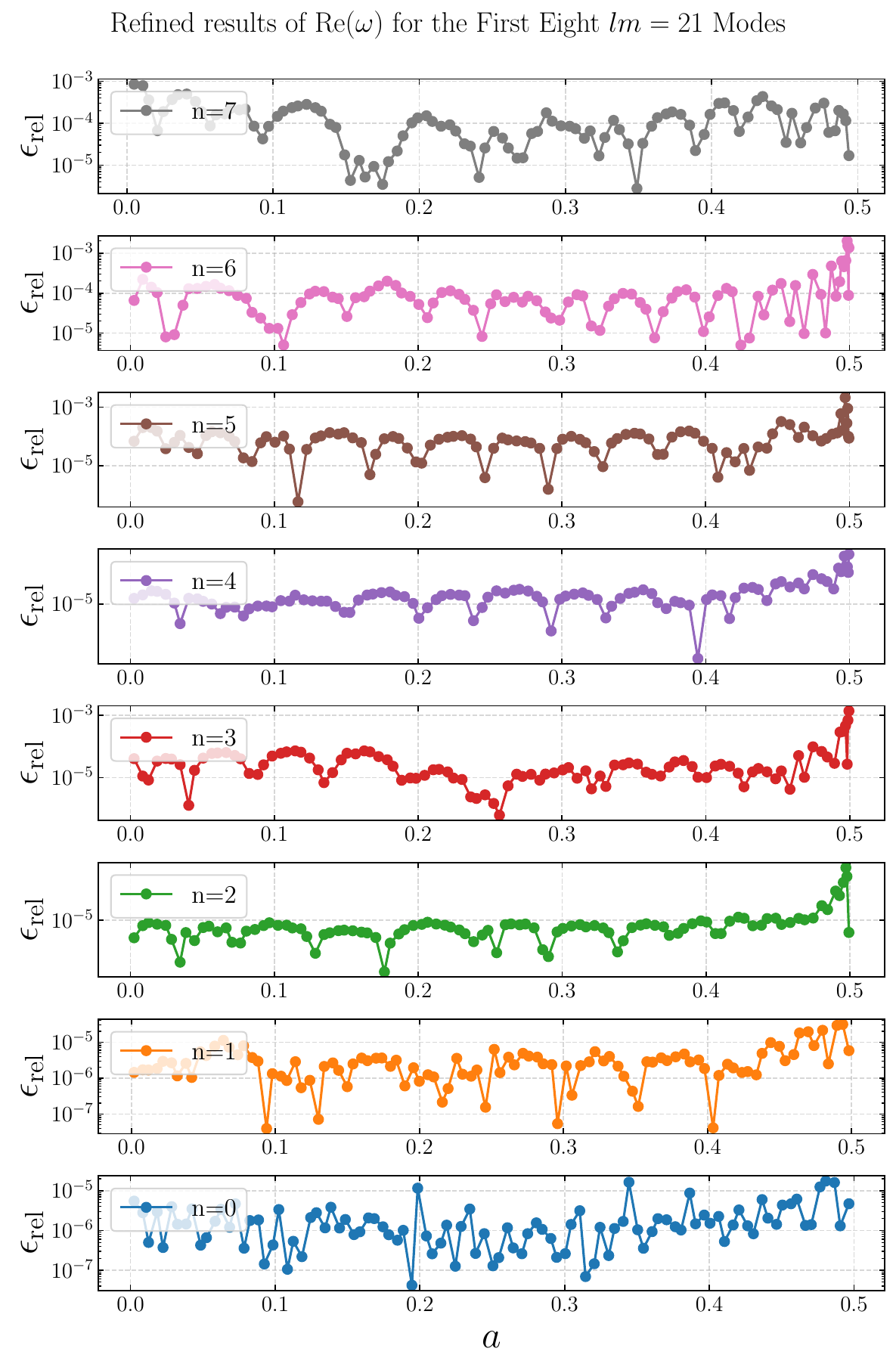}
    \end{minipage}
    \caption{Solution accuracy for the real part of the first eight QNMs in the $lm=21$ case obtained with \texttt{DeepOPiraKAN}, illustrating the effect of evaluation resolution. The left panel shows results obtained using $2\times10^3$ evaluation steps, while the right panel shows results obtained using $4\times10^3$ evaluation steps.}
    \label{lm21_Re} 	
\end{figure*}

\begin{figure*}[htbp]
\centering
    \begin{minipage}{0.48\textwidth}
        \centering
        \includegraphics[width=\textwidth]{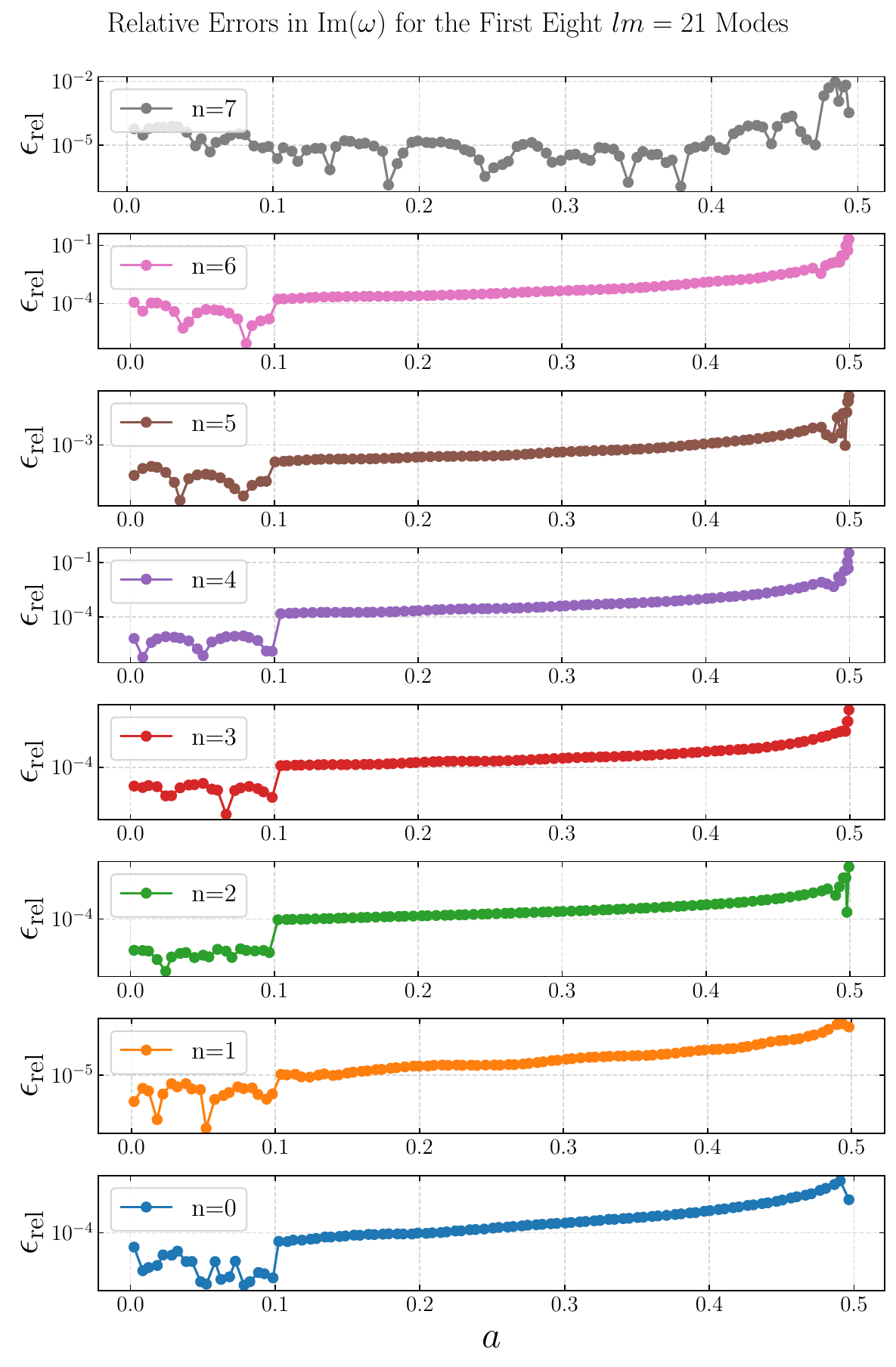}
    \end{minipage}
    \hspace{0.01\textwidth}
    \begin{minipage}{0.48\textwidth}
        \centering
        \includegraphics[width=\textwidth]{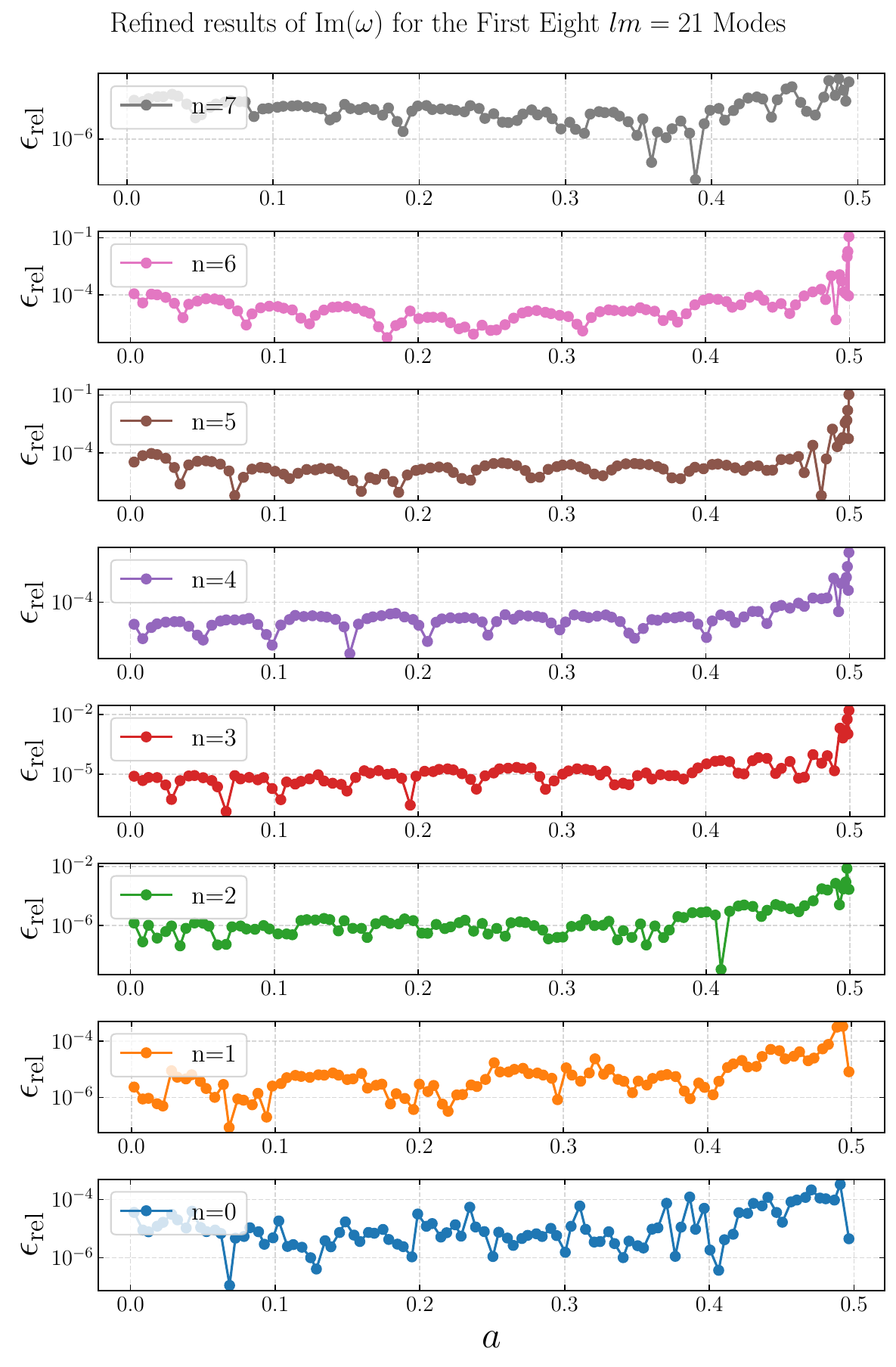}
    \end{minipage}
    \caption{Solution accuracy for the imaginary part of the first eight QNMs in the $lm=21$ case, shown under the same conditions as Fig.~\ref{lm21_Re}.}
    \label{lm21_Im} 	
\end{figure*}

The results for the first eight modes with $lm=20$ are shown in Fig.~\ref{lm20}, obtained using $2\times10^3$ evaluation steps. The left panel displays the relative error in the real part of the eigenfrequency $\omega$ with respect to the benchmark values computed using the Leaver's method, while the right panel shows the corresponding results for the imaginary part. The relative error is defined as
\begin{equation}
\epsilon_{\rm rel} = \left|\frac{\omega_{\rm PINN} - \omega_{\rm Leaver}}{\omega_{\rm Leaver}}\right|,
\end{equation}
where $\omega_{\rm Leaver}$ is obtained using the \texttt{qnm} package~\cite{Stein:2019mop}. In each panel, from bottom to top, the curves correspond to successive mode indices as functions of the spin parameter $a \in [0,0.5)$.
We find that, in the low overtone regime, the relative errors in $\mathrm{Re}(\omega)$ and $\mathrm{Im}(\omega)$ are of comparable magnitude. As the overtone index increases, the relative error in $\mathrm{Im}(\omega)$ becomes systematically smaller than that in $\mathrm{Re}(\omega)$, primarily due to the rapid growth of $|\mathrm{Im}(\omega)|$ with increasing $n$ \cite{Leaver:1985ax}. Overall, the relative errors for both components exhibit a clear and systematic scaling with overtone number. For example, the error in $\mathrm{Re}(\omega)$ increases from $\mathcal{O}(10^{-5})$ for the fundamental mode to $\mathcal{O}(10^{-3})$ at $n=7$, with $\mathrm{Im}(\omega)$ following a similar trend. Importantly, for a fixed mode index, the error magnitude remains largely insensitive to variations in the spin parameter $a$, showing only mild fluctuations within the 
similar order of magnitude. Together, these features indicate that the learned spectral mapping preserves the intrinsic structure of the QNM spectrum across continuous parameter space, capturing key aspects of its hierarchical organization through a stable and well conditioned operator representation, rather than merely interpolating isolated eigenvalues.

The real part of the first eight QNM modes with $lm=21$ is shown in Fig.~\ref{lm21_Re}. The left panel presents results obtained using $2\times10^3$ evaluation steps. Overall, the error behavior closely follows that observed for the $lm=20$ modes, with relative errors increasing systematically with overtone number. A notable deviation appears for the first seven modes at spin values $a>0.1$, where the relative error exhibits a pronounced increase. This behavior originates from the spin–mode coupling term in Eq.~(\ref{radial_part}), which becomes active for $m=1$ and introduces increased numerical stiffness at higher spins.
Importantly, this degradation does not indicate a breakdown of the learned operator representation, but rather reflects the increased resolution required to accurately resolve the spectrum in this strongly coupled regime. As shown in the right panel of Fig.~\ref{lm21_Re}, increasing the number of evaluation steps to $4\times10^3$ restores smooth convergence and removes the discontinuity. With this refinement, the relative error in $\mathrm{Re}(\omega)$ ranges from $\mathcal{O}(10^{-6})$ for the fundamental mode to $\mathcal{O}(10^{-4})$ at $n=7$. The imaginary part exhibits consistent behavior. When sufficient evaluation resolution is employed, the relative error in $\mathrm{Im}(\omega)$ follows the same systematic scaling with overtone number and remains well controlled across the full spin range, as shown in Fig.~\ref{lm21_Im}.
The $lm=21$ modes provide a stringent stress test for the proposed architecture, demonstrating that \texttt{DeepOPiraKAN} maintains stable and systematically controllable accuracy even in the presence of strong spin–mode coupling, without requiring retraining or mode specific adjustments. 

Taken together, these results establish \texttt{DeepOPiraKAN} as both a new physics informed operator architecture and a practical framework for parameter dependent spectral computation. A single trained model resolves the Kerr QNM spectrum across the full spin range, captures both fundamental and overtone structure in the $(\ell,m)=(2,0)$ and $(2,1)$ families, and maintains systematically controllable accuracy even in the strongly coupled regime without mode-specific retraining. Our results point to a practically viable route toward high-precision QNM template infrastructures, demonstrating that neural operator learning can transform the computation of Kerr spectra from a pointwise numerical task into a scalable surrogate problem, with a level of accuracy that is promising for future black hole spectroscopy~\cite{Carullo:2025oms}. More broadly, the architecture provides a promising template for other parameter-dependent spectral problems in which complex spectra must be learned as continuous functions of physical parameters.

\section{Discussion and Summary}
\label{summary}
\label{summary}
\begin{figure*}[htbp]
\centering
    \begin{minipage}{0.48\textwidth}
        \centering
        \includegraphics[width=\textwidth]{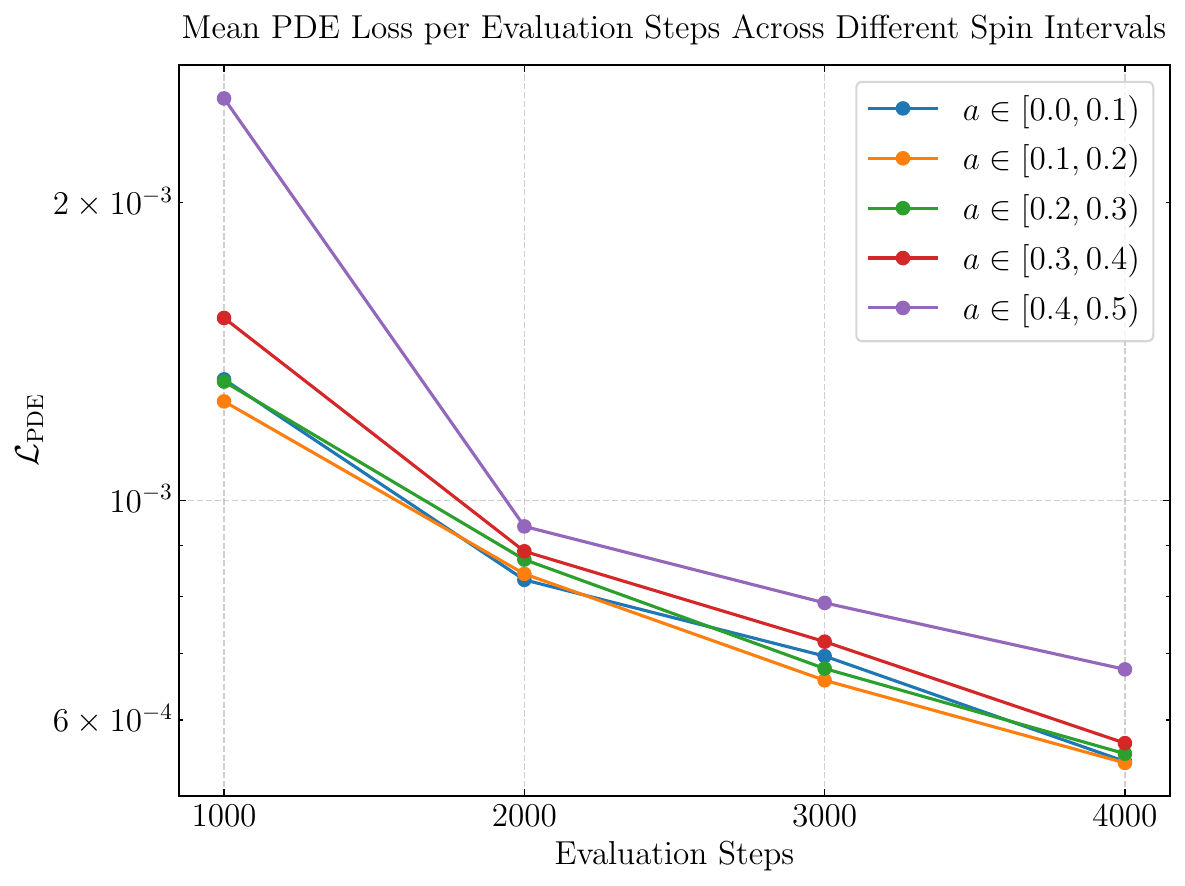}
    \end{minipage}
    \hspace{0.01\textwidth}
    \begin{minipage}{0.48\textwidth}
        \centering
        \includegraphics[width=\textwidth]{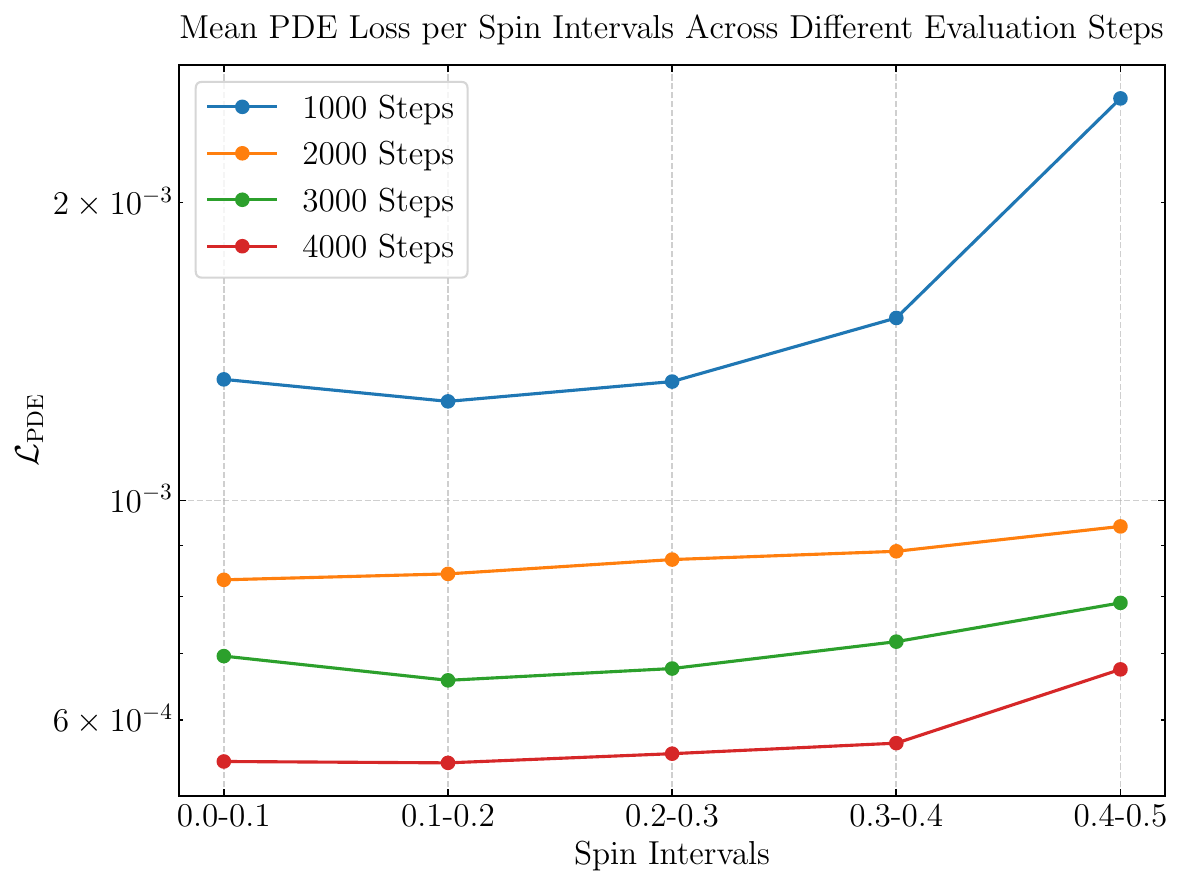}
    \end{minipage}
    \caption{Mean PDE loss $\mathcal{L}_{\mathrm{PDE}}$ for the $lmn=211$ mode as a function of spin interval and evaluation resolution. The left panel shows the dependence of $\mathcal{L}_{\mathrm{PDE}}$ on the number of evaluation steps for different spin intervals, while the right panel displays its variation with spin interval at fixed evaluation steps.}
    \label{ped_loss} 	
\end{figure*}
Beyond the benchmark performance reported above, the present results also reveal an important methodological feature of \texttt{DeepOPiraKAN}. Its predictive accuracy is controlled not only by training, but also by the resolution at which the learned operator is evaluated during inference. For the $lm=21$ mode family, using $2\times10^3$ evaluation steps leads to visible degradation of accuracy at higher spins ($a>0.1$), whereas increasing the evaluation steps to $4\times10^3$ restores smooth convergence across the full parameter range. As shown in Fig.~\ref{ped_loss}, the mean PDE loss $\mathcal{L}_{\mathrm{PDE}}$ decreases systematically with increasing evaluation resolution, with the most pronounced reduction occurring between $10^3$ and $2\times10^3$ steps, followed by a more gradual improvement up to $4\times10^3$ steps. For a fixed evaluation budget, higher-spin configurations yield larger values of $\mathcal{L}_{\mathrm{PDE}}$, consistent with the increased stiffness of the underlying equations. This behaviour indicates that the learned mapping is globally constrained and can be systematically refined at inference time without retraining. Such a separation between operator learning and evaluation resolution is a distinctive advantage over pointwise eigensolvers, enabling predictable and controllable accuracy across continuous parameter space.

More broadly, beyond the specific Kerr benchmark studied here, \texttt{DeepOPiraKAN} provides a general operator learning framework for parameter dependent spectral problems governed by differential operators. At the methodological level, its modular design allows alternative functional representations, including other orthogonal polynomial bases or wavelet expansions, to be incorporated without changing the core learning strategy. At the application level, a particularly natural next application is to neutron star oscillation spectra with realistic equations of state, where spectral calculations remain highly challenging because the modes depend sensitively on multiple continuous parameters and conventional mode-by-mode solvers are computationally costly~\cite{Kokkotas:1992ka,Benhar:1998au}. Furthermore, this framework suggests a scalable route for spectral computation across physics and engineering, especially in systems where complex spectra depend nontrivially on continuous parameters. Together with the open-source implementation released in this work, our results position learning based neural operators as a promising new paradigm for complex spectral analysis.

\emph{\textbf{Acknowledgements}.---}
We thank Sifan Wang and Huan Yang for helpful discussion and comments. This work is supported by National Natural Science Foundation of China under Grants No.~12335006. ZL is supported by ``the Fundamental Research Funds for the Central Universities''. It is also supported by the High-performance Computing Platform of Peking University.

\bibliography{scibib}

@misc{wang2020understandingmitigatinggradientpathologies,
      title={Understanding and mitigating gradient pathologies in physics-informed neural networks}, 
      author={Sifan Wang and Yujun Teng and Paris Perdikaris},
      year={2020},
      eprint={2001.04536},
      archivePrefix={arXiv},
      primaryClass={cs.LG},
      url={https://arxiv.org/abs/2001.04536}, 
}

@misc{anagnostopoulos2023residualbasedattentionconnectioninformation,
      title={Residual-based attention and connection to information bottleneck theory in PINNs}, 
      author={Sokratis J. Anagnostopoulos and Juan Diego Toscano and Nikolaos Stergiopulos and George Em Karniadakis},
      year={2023},
      eprint={2307.00379},
      archivePrefix={arXiv},
      primaryClass={cs.LG},
      url={https://arxiv.org/abs/2307.00379}, 
}

@article{Wang_2021,
   title={On the eigenvector bias of Fourier feature networks: From regression to solving multi-scale PDEs with physics-informed neural networks},
   volume={384},
   ISSN={0045-7825},
   url={http://dx.doi.org/10.1016/j.cma.2021.113938},
   DOI={10.1016/j.cma.2021.113938},
   journal={Computer Methods in Applied Mechanics and Engineering},
   publisher={Elsevier BV},
   author={Wang, Sifan and Wang, Hanwen and Perdikaris, Paris},
   year={2021},
   month=oct, pages={113938} }

@misc{tancik2020fourierfeaturesletnetworks,
      title={Fourier Features Let Networks Learn High Frequency Functions in Low Dimensional Domains}, 
      author={Matthew Tancik and Pratul P. Srinivasan and Ben Mildenhall and Sara Fridovich-Keil and Nithin Raghavan and Utkarsh Singhal and Ravi Ramamoorthi and Jonathan T. Barron and Ren Ng},
      year={2020},
      eprint={2006.10739},
      archivePrefix={arXiv},
      primaryClass={cs.CV},
      url={https://arxiv.org/abs/2006.10739}, 
}

@misc{wang2024piratenetsphysicsinformeddeeplearning,
      title={PirateNets: Physics-informed Deep Learning with Residual Adaptive Networks}, 
      author={Sifan Wang and Bowen Li and Yuhan Chen and Paris Perdikaris},
      year={2024},
      eprint={2402.00326},
      archivePrefix={arXiv},
      primaryClass={cs.LG},
      url={https://arxiv.org/abs/2402.00326}, 
}

@misc{liu2025kankolmogorovarnoldnetworks,
      title={KAN: Kolmogorov-Arnold Networks}, 
      author={Ziming Liu and Yixuan Wang and Sachin Vaidya and Fabian Ruehle and James Halverson and Marin Soljačić and Thomas Y. Hou and Max Tegmark},
      year={2025},
      eprint={2404.19756},
      archivePrefix={arXiv},
      primaryClass={cs.LG},
      url={https://arxiv.org/abs/2404.19756}, 
}

@article{Wang_2025,
   title={Kolmogorov–Arnold-Informed neural network: A physics-informed deep learning framework for solving forward and inverse problems based on Kolmogorov–Arnold Networks},
   volume={433},
   ISSN={0045-7825},
   url={http://dx.doi.org/10.1016/j.cma.2024.117518},
   DOI={10.1016/j.cma.2024.117518},
   journal={Computer Methods in Applied Mechanics and Engineering},
   publisher={Elsevier BV},
   author={Wang, Yizheng and Sun, Jia and Bai, Jinshuai and Anitescu, Cosmin and Eshaghi, Mohammad Sadegh and Zhuang, Xiaoying and Rabczuk, Timon and Liu, Yinghua},
   year={2025},
   month=jan, pages={117518} }

@article{Hu:2017mde,
    author = "Hu, Wen-Rui and Wu, Yue-Liang",
    title = "{The Taiji Program in Space for gravitational wave physics and the nature of gravity}",
    doi = "10.1093/nsr/nwx116",
    journal = "Natl. Sci. Rev.",
    volume = "4",
    number = "5",
    pages = "685--686",
    year = "2017"
}

@misc{zhang2025kolmogorovarnoldfouriernetworks,
      title={Kolmogorov-Arnold Fourier Networks}, 
      author={Jusheng Zhang and Yijia Fan and Kaitong Cai and Keze Wang},
      year={2025},
      eprint={2502.06018},
      archivePrefix={arXiv},
      primaryClass={cs.LG},
      url={https://arxiv.org/abs/2502.06018}, 
}

@misc{farea2025learnableactivationfunctionsphysicsinformed,
      title={Learnable Activation Functions in Physics-Informed Neural Networks for Solving Partial Differential Equations}, 
      author={Afrah Farea and Mustafa Serdar Celebi},
      year={2025},
      eprint={2411.15111},
      archivePrefix={arXiv},
      primaryClass={cs.NE},
      url={https://arxiv.org/abs/2411.15111}, 
}

@article{Lu_2021,
   title={Learning nonlinear operators via DeepONet based on the universal approximation theorem of operators},
   volume={3},
   ISSN={2522-5839},
   url={http://dx.doi.org/10.1038/s42256-021-00302-5},
   DOI={10.1038/s42256-021-00302-5},
   number={3},
   journal={Nature Machine Intelligence},
   publisher={Springer Science and Business Media LLC},
   author={Lu, Lu and Jin, Pengzhan and Pang, Guofei and Zhang, Zhongqiang and Karniadakis, George Em},
   year={2021},}

@misc{jin_2022,
  title={MIONet: Learning multiple-input operators via tensor product}, 
  author={Pengzhan Jin and Shuai Meng and Lu Lu},
  year={2022},
  eprint={2202.06137},
  archivePrefix={arXiv},
  primaryClass={cs.LG},
  url={https://arxiv.org/abs/2202.06137}, 
}

@article{
doi:10.1126/sciadv.abi8605,
author = {Sifan Wang  and Hanwen Wang  and Paris Perdikaris },
title = {Learning the solution operator of parametric partial differential equations with physics-informed DeepONets},
journal = {Science Advances},
volume = {7},
number = {40},
pages = {eabi8605},
year = {2021},
doi = {10.1126/sciadv.abi8605},
URL = {https://www.science.org/doi/abs/10.1126/sciadv.abi8605},
eprint = {https://www.science.org/doi/pdf/10.1126/sciadv.abi8605},}

@article{Kokkotas:1999bd,
    author = "Kokkotas, Kostas D. and Schmidt, Bernd G.",
    title = "{Quasinormal modes of stars and black holes}",
    eprint = "gr-qc/9909058",
    archivePrefix = "arXiv",
    doi = "10.12942/lrr-1999-2",
    journal = "Living Rev. Rel.",
    volume = "2",
    pages = "2",
    year = "1999"
}

@article{LIGOScientific:2016aoc,
    author = "Abbott, B. P. and others",
    collaboration = "LIGO Scientific, Virgo",
    title = "{Observation of Gravitational Waves from a Binary Black Hole Merger}",
    eprint = "1602.03837",
    archivePrefix = "arXiv",
    primaryClass = "gr-qc",
    reportNumber = "LIGO-P150914",
    doi = "10.1103/PhysRevLett.116.061102",
    journal = "Phys. Rev. Lett.",
    volume = "116",
    number = "6",
    pages = "061102",
    year = "2016"
}

@article{LIGOScientific:2016lio,
    author = "Abbott, B. P. and others",
    collaboration = "LIGO Scientific, Virgo",
    title = "{Tests of general relativity with GW150914}",
    eprint = "1602.03841",
    archivePrefix = "arXiv",
    primaryClass = "gr-qc",
    reportNumber = "LIGO-P1500213",
    doi = "10.1103/PhysRevLett.116.221101",
    journal = "Phys. Rev. Lett.",
    volume = "116",
    number = "22",
    pages = "221101",
    year = "2016",
    note = "[Erratum: Phys.Rev.Lett. 121, 129902 (2018)]"
}

@article{LIGOScientific:2025rsn,
    author = "Abac, A. G. and others",
    collaboration = "LIGO Scientific, VIRGO, KAGRA",
    title = "{GW231123: A Binary Black Hole Merger with Total Mass 190{\textendash}265 M$_{\odot}$}",
    eprint = "2507.08219",
    archivePrefix = "arXiv",
    primaryClass = "astro-ph.HE",
    reportNumber = "DCC: P2500026-v6",
    doi = "10.3847/2041-8213/ae0c9c",
    journal = "Astrophys. J. Lett.",
    volume = "993",
    number = "1",
    pages = "L25",
    year = "2025"
}

@article{LIGOScientific:2025obp,
    collaboration = "LIGO Scientific, VIRGO, KAGRA",
    title = "{Black Hole Spectroscopy and Tests of General Relativity with GW250114}",
    eprint = "2509.08099",
    archivePrefix = "arXiv",
    primaryClass = "gr-qc",
    reportNumber = "LIGO P2500461",
    month = "9",
    year = "2025"
}

@article{Bhagwat:2021kwv,
    author = "Bhagwat, Swetha and Pacilio, Costantino and Barausse, Enrico and Pani, Paolo",
    title = "{Landscape of massive black-hole spectroscopy with LISA and the Einstein Telescope}",
    eprint = "2201.00023",
    archivePrefix = "arXiv",
    primaryClass = "gr-qc",
    reportNumber = "ET-0465A-21",
    doi = "10.1103/PhysRevD.105.124063",
    journal = "Phys. Rev. D",
    volume = "105",
    number = "12",
    pages = "124063",
    year = "2022"
}

@article{Berti:2007fi,
    author = "Berti, Emanuele and Cardoso, Vitor and Gonzalez, Jose A. and Sperhake, Ulrich and Hannam, Mark and Husa, Sascha and Bruegmann, Bernd",
    title = "{Inspiral, merger and ringdown of unequal mass black hole binaries: A Multipolar analysis}",
    eprint = "gr-qc/0703053",
    archivePrefix = "arXiv",
    doi = "10.1103/PhysRevD.76.064034",
    journal = "Phys. Rev. D",
    volume = "76",
    pages = "064034",
    year = "2007"
}

@article{Teukolsky:1973ha,
    author = "Teukolsky, Saul A.",
    title = "{Perturbations of a rotating black hole. 1. Fundamental equations for gravitational electromagnetic and neutrino field perturbations}",
    doi = "10.1086/152444",
    journal = "Astrophys. J.",
    volume = "185",
    pages = "635--647",
    year = "1973"
}

@article{Teukolsky:1972my,
    author = "Teukolsky, S. A.",
    title = "{Rotating black holes - separable wave equations for gravitational and electromagnetic perturbations}",
    reportNumber = "OAP-291",
    doi = "10.1103/PhysRevLett.29.1114",
    journal = "Phys. Rev. Lett.",
    volume = "29",
    pages = "1114--1118",
    year = "1972"
}

@article{Leaver:1985ax,
    author = "Leaver, E. W.",
    title = "{An Analytic representation for the quasi normal modes of Kerr black holes}",
    doi = "10.1098/rspa.1985.0119",
    journal = "Proc. Roy. Soc. Lond. A",
    volume = "402",
    pages = "285--298",
    year = "1985"
}

@article{Vishveshwara:1970zz,
    author = "Vishveshwara, C. V.",
    title = "{Scattering of Gravitational Radiation by a Schwarzschild Black-hole}",
    doi = "10.1038/227936a0",
    journal = "Nature",
    volume = "227",
    pages = "936--938",
    year = "1970"
}

@article{Chandrasekhar:1975zza,
    author = "Chandrasekhar, S. and Detweiler, Steven L.",
    title = "{The quasi-normal modes of the Schwarzschild black hole}",
    doi = "10.1098/rspa.1975.0112",
    journal = "Proc. Roy. Soc. Lond. A",
    volume = "344",
    pages = "441--452",
    year = "1975"
}

@article{Schutz:1985km,
    author = "Schutz, Bernard F. and Will, Clifford M.",
    title = "{BLACK HOLE NORMAL MODES: A SEMIANALYTIC APPROACH}",
    reportNumber = "PRINT-85-0063 (WASH.U.,ST.LOUIS)",
    doi = "10.1086/184453",
    journal = "Astrophys. J. Lett.",
    volume = "291",
    pages = "L33--L36",
    year = "1985"
}

@article{Iyer:1986np,
    author = "Iyer, Sai and Will, Clifford M.",
    title = "{Black Hole Normal Modes: A {WKB} Approach. 1. Foundations and Application of a Higher Order {WKB} Analysis of Potential Barrier Scattering}",
    reportNumber = "Print-86-1482 (WASH. U., ST. LOUIS)",
    doi = "10.1103/PhysRevD.35.3621",
    journal = "Phys. Rev. D",
    volume = "35",
    pages = "3621",
    year = "1987"
}

@article{Raissi:2017zsi,
    author = "Raissi, Maziar and Perdikaris, Paris and Karniadakis, George Em",
    title = "{Physics-informed neural networks: A deep learning framework for solving forward and inverse problems involving nonlinear partial differential equations}",
    eprint = "1711.10561",
    archivePrefix = "arXiv",
    primaryClass = "cs.AI",
    doi = "10.1016/j.jcp.2018.10.045",
    journal = "J. Comput. Phys.",
    volume = "378",
    pages = "686--707",
    year = "2019"
}

@article{Luna:2024spo,
    author = "Luna, Raimon and Doneva, Daniela D. and Font, Jos{\'e} A. and Lien, Jr-Hua and Yazadjiev, Stoytcho S.",
    title = "{Quasinormal modes in modified gravity using physics-informed neural networks}",
    eprint = "2404.11583",
    archivePrefix = "arXiv",
    primaryClass = "gr-qc",
    doi = "10.1103/PhysRevD.109.124064",
    journal = "Phys. Rev. D",
    volume = "109",
    number = "12",
    pages = "124064",
    year = "2024"
}

@article{Luna:2022rql,
    author = "Luna, Raimon and Calder{\'o}n Bustillo, Juan and Mart{\'\i}nez, Juan Jos{\'e} Seoane and Torres-Forn{\'e}, Alejandro and Font, Jos{\'e} A.",
    title = "{Solving the Teukolsky equation with physics-informed neural networks}",
    eprint = "2212.06103",
    archivePrefix = "arXiv",
    primaryClass = "gr-qc",
    doi = "10.1103/PhysRevD.107.064025",
    journal = "Phys. Rev. D",
    volume = "107",
    number = "6",
    pages = "064025",
    year = "2023"
}

@article{Cornell:2024azz,
    author = "Cornell, Alan S. and Herbst, Sheldon R. and Ncube, Anele M. and Noshad, Hajar",
    title = "{Solving the Regge-Wheeler and Teukolsky equations: supervised versus unsupervised physics-informed neural networks}",
    eprint = "2402.11343",
    archivePrefix = "arXiv",
    primaryClass = "gr-qc",
    month = "2",
    year = "2024"
}

@article{Cornell:2022enn,
    author = "Cornell, Alan S. and Ncube, Anele and Harmsen, Gerhard",
    title = "{Using physics-informed neural networks to compute quasinormal modes}",
    eprint = "2205.08284",
    archivePrefix = "arXiv",
    primaryClass = "physics.comp-ph",
    doi = "10.1103/PhysRevD.106.124047",
    journal = "Phys. Rev. D",
    volume = "106",
    number = "12",
    pages = "124047",
    year = "2022"
}

@article{Patel:2024wzo,
    author = "Patel, Nirmal and Aykutalp, Aycin and Laguna, Pablo",
    title = "{Calculating Quasi-Normal Modes of Schwarzschild Black Holes with Physics Informed Neural Networks}",
    eprint = "2401.01440",
    archivePrefix = "arXiv",
    primaryClass = "gr-qc",
    month = "1",
    year = "2024"
}

@article{Pombo:2025urp,
    author = "Pombo, Alexandre M. and Pizzuti, Lorenzo",
    title = "{Teukolsky by Design: A Hybrid Spectral-PINN solver for Kerr Quasinormal Modes}",
    eprint = "2511.15796",
    archivePrefix = "arXiv",
    primaryClass = "gr-qc",
    month = "11",
    year = "2025"
}

@article{Konoplya:2011qq,
    author = "Konoplya, R. A. and Zhidenko, A.",
    title = "{Quasinormal modes of black holes: From astrophysics to string theory}",
    eprint = "1102.4014",
    archivePrefix = "arXiv",
    primaryClass = "gr-qc",
    doi = "10.1103/RevModPhys.83.793",
    journal = "Rev. Mod. Phys.",
    volume = "83",
    pages = "793--836",
    year = "2011"
}

@article{Berti:2009kk,
    author = "Berti, Emanuele and Cardoso, Vitor and Starinets, Andrei O.",
    title = "{Quasinormal modes of black holes and black branes}",
    eprint = "0905.2975",
    archivePrefix = "arXiv",
    primaryClass = "gr-qc",
    doi = "10.1088/0264-9381/26/16/163001",
    journal = "Class. Quant. Grav.",
    volume = "26",
    pages = "163001",
    year = "2009"
}

@incollection{anitescu_2023,
  title={Physics-informed neural networks: Theory and applications},
  author={Anitescu, Cosmin and {\.I}smail Ate{\c{s}}, Burak and Rabczuk, Timon},
  booktitle={Machine Learning in Modeling and Simulation: Methods and Applications},
  pages={179--218},
  year={2023},
  publisher={Springer}
}

@misc{wang_soap_2025,
      title={Gradient Alignment in Physics-informed Neural Networks: A Second-Order Optimization Perspective}, 
      author={Sifan Wang and Ananyae Kumar Bhartari and Bowen Li and Paris Perdikaris},
      year={2025},
      eprint={2502.00604},
      archivePrefix={arXiv},
      primaryClass={cs.LG},
      url={https://arxiv.org/abs/2502.00604}, 
}

@InProceedings{pmlr-v202-daw23a,
  title = 	 {Mitigating Propagation Failures in Physics-informed Neural Networks using Retain-Resample-Release ({R}3) Sampling},
  author =       {Daw, Arka and Bu, Jie and Wang, Sifan and Perdikaris, Paris and Karpatne, Anuj},
  booktitle = 	 {Proceedings of the 40th International Conference on Machine Learning},
  pages = 	 {7264--7302},
  year = 	 {2023},
  editor = 	 {Krause, Andreas and Brunskill, Emma and Cho, Kyunghyun and Engelhardt, Barbara and Sabato, Sivan and Scarlett, Jonathan},
  volume = 	 {202},
  series = 	 {Proceedings of Machine Learning Research},
  month = 	 {23--29 Jul},
  publisher =    {PMLR},
  url = 	 {https://proceedings.mlr.press/v202/daw23a.html},
}

@misc{rathore2024challengestrainingpinnsloss,
      title={Challenges in Training PINNs: A Loss Landscape Perspective}, 
      author={Pratik Rathore and Weimu Lei and Zachary Frangella and Lu Lu and Madeleine Udell},
      year={2024},
      eprint={2402.01868},
      archivePrefix={arXiv},
      primaryClass={cs.LG},
      url={https://arxiv.org/abs/2402.01868}, 
}

@article{GOPAKUMAR2023100464,
title = {Loss landscape engineering via Data Regulation on PINNs},
journal = {Machine Learning with Applications},
volume = {12},
pages = {100464},
year = {2023},
issn = {2666-8270},
doi = {https://doi.org/10.1016/j.mlwa.2023.100464},
url = {https://www.sciencedirect.com/science/article/pii/S2666827023000178},
author = {Vignesh Gopakumar and Stanislas Pamela and Debasmita Samaddar},
}

@article{Yang:2012he,
    author = "Yang, Huan and Nichols, David A. and Zhang, Fan and Zimmerman, Aaron and Zhang, Zhongyang and Chen, Yanbei",
    title = "{Quasinormal-mode spectrum of Kerr black holes and its geometric interpretation}",
    eprint = "1207.4253",
    archivePrefix = "arXiv",
    primaryClass = "gr-qc",
    doi = "10.1103/PhysRevD.86.104006",
    journal = "Phys. Rev. D",
    volume = "86",
    pages = "104006",
    year = "2012"
}

@article{Ferrari:1984zz,
    author = "Ferrari, Valeria and Mashhoon, Bahram",
    title = "{New approach to the quasinormal modes of a black hole}",
    doi = "10.1103/PhysRevD.30.295",
    journal = "Phys. Rev. D",
    volume = "30",
    pages = "295--304",
    year = "1984"
}

@software{QuasiNormalModesToolkit,
  author       = {O'Toole, Conor and Macedo, Rodrigo and Stratton, Tom and Wardell, Barry},
  title        = {QuasiNormalModes: A Black Hole Perturbation Toolkit Package},
  year         = {2025},
  publisher    = {GitHub},
  howpublished = {\url{https://github.com/BlackHolePerturbationToolkit/QuasiNormalModes}},
  note         = {A package for computing quasinormal modes of Schwarzschild and Kerr black holes}
}

@article{Stein:2019mop,
    author = "Stein, Leo C.",
    title = "{qnm: A Python package for calculating Kerr quasinormal modes, separation constants, and spherical-spheroidal mixing coefficients}",
    eprint = "1908.10377",
    archivePrefix = "arXiv",
    primaryClass = "gr-qc",
    doi = "10.21105/joss.01683",
    journal = "J. Open Source Softw.",
    volume = "4",
    number = "42",
    pages = "1683",
    year = "2019"
}

@article{Tseneklidou:2025stn,
    author = "Tseneklidou, Dimitra and Torres-Forne, Alejandro and Cerda-Duran, Pablo",
    title = "{Towards asteroseismology of neutron stars with physics-informed neural networks}",
    eprint = "2504.12183",
    archivePrefix = "arXiv",
    primaryClass = "astro-ph.HE",
    doi = "10.1140/epjc/s10052-025-14942-z",
    journal = "Eur. Phys. J. C",
    volume = "85",
    number = "10",
    pages = "1218",
    year = "2025"
}

@Misc{paddle,
howpublished = {\url{https://www.paddlepaddle.org.cn/en}},
year = {2025},
author = "Baidu Inc",
}

@Misc{ppsci,
howpublished = {\url{https://github.com/PaddlePaddle/PaddleScience/tree/develop/examples/qnm}},
year = {2025},
}

@article{LISA:2017pwj,
    author = "Amaro-Seoane, Pau and others",
    collaboration = "LISA",
    title = "{Laser Interferometer Space Antenna}",
    journal = {e-prints},
    eprint = "1702.00786",
    archivePrefix = "arXiv",
    primaryClass = "astro-ph.IM",
    month = "2",
    year = "2017"
}

@article{TianQin:2015yph,
    author = "Luo, Jun and others",
    collaboration = "TianQin",
    title = "{TianQin: a space-borne gravitational wave detector}",
    eprint = "1512.02076",
    archivePrefix = "arXiv",
    primaryClass = "astro-ph.IM",
    doi = "10.1088/0264-9381/33/3/035010",
    journal = "Class. Quant. Grav.",
    volume = "33",
    number = "3",
    pages = "035010",
    year = "2016"
}

@article{Du:2025xdq,
    author = "Du, Minghui and others",
    title = "{Towards realistic detection pipelines of Taiji: New challenges in data analysis and high-fidelity simulations of space-based gravitational wave antenna}",
    eprint = "2505.16500",
    archivePrefix = "arXiv",
    primaryClass = "gr-qc",
    doi = "10.1007/s11433-025-2870-8",
    journal = "Sci. China Phys. Mech. Astron.",
    volume = "69",
    number = "4",
    pages = "249501",
    year = "2026"
}

@article{Reitze:2019iox,
	title        = {{Cosmic Explorer: The U.S. Contribution to Gravitational-Wave Astronomy beyond LIGO}},
	author       = {Reitze, David and others},
	year         = 2019,
	journal      = {Bull. Am. Astron. Soc.},
	volume       = 51,
	number       = 7,
	pages        = {035},
	eprint       = {1907.04833},
	archiveprefix = {arXiv},
	primaryclass = {astro-ph.IM},
	reportnumber = {LIGO-P1900316}
}

@article{Punturo:2010zz,
	title        = {{The Einstein Telescope: A third-generation gravitational wave observatory}},
	author       = {Punturo, M. and others},
	year         = 2010,
	journal      = {Class. Quant. Grav.},
	volume       = 27,
	pages        = 194002,
	doi          = {10.1088/0264-9381/27/19/194002},
	editor       = {Ricci, Fulvio}
}

@inproceedings{hernandez2002slepc,
  title={SLEPc: Scalable library for eigenvalue problem computations},
  author={Hern{\'a}ndez, Vicente and Rom{\'a}n, Jose E and Vidal, Vicente},
  booktitle={International Conference on High Performance Computing for Computational Science},
  pages={377--391},
  year={2002},
  organization={Springer}
}

@article{roman2016slepc,
  title={SLEPc users manual},
  author={Roman, Jose E and Campos, Carmen and Romero, Eloy and Tom{\'a}s, Andr{\'e}s},
  journal={Departamento di Sistemas Informaticos y Computaci{\'o}n, Universitat Politecnica de Valencia, TR DSIC-II/24/02, Rev},
  volume={3},
  year={2016}
}

@inbook{Franchini:2023eda,
    author = {Franchini, Nicola and V{\"o}lkel, Sebastian H.},
    title = "{Testing General Relativity with Black Hole Quasi-normal Modes}",
    eprint = "2305.01696",
    archivePrefix = "arXiv",
    primaryClass = "gr-qc",
    doi = "10.1007/978-981-97-2871-8_9",
    year = "2024"
}

@article{Berti:2025hly,
    author = "Abedi, Jahed and others",
    editor = "Berti, Emanuele and Cardoso, Vitor and Carullo, Gregorio",
    title = "{Black hole spectroscopy: from theory to experiment}",
    eprint = "2505.23895",
    archivePrefix = "arXiv",
    primaryClass = "gr-qc",
    month = "5",
    year = "2025"
}

@article{Barausse:2014tra,
    author = "Barausse, Enrico and Cardoso, Vitor and Pani, Paolo",
    title = "{Can environmental effects spoil precision gravitational-wave astrophysics?}",
    eprint = "1404.7149",
    archivePrefix = "arXiv",
    primaryClass = "gr-qc",
    doi = "10.1103/PhysRevD.89.104059",
    journal = "Phys. Rev. D",
    volume = "89",
    number = "10",
    pages = "104059",
    year = "2014"
}

@article{Baibhav:2018rfk,
    author = "Baibhav, Vishal and Berti, Emanuele",
    title = "{Multimode black hole spectroscopy}",
    eprint = "1809.03500",
    archivePrefix = "arXiv",
    primaryClass = "gr-qc",
    doi = "10.1103/PhysRevD.99.024005",
    journal = "Phys. Rev. D",
    volume = "99",
    number = "2",
    pages = "024005",
    year = "2019"
}

@article{Carullo:2025oms,
    author = "Carullo, Gregorio",
    title = "{Black hole spectroscopy: status report}",
    doi = "10.1007/s10714-025-03408-y",
    journal = "Gen. Rel. Grav.",
    volume = "57",
    number = "5",
    pages = "76",
    year = "2025"
}

@article{Kokkotas:1992ka,
    author = "Kokkotas, K. D. and Schutz, Bernard F.",
    title = "{W-modes: A New family of normal modes of pulsating relativistic stars}",
    journal = "Mon. Not. Roy. Astron. Soc.",
    volume = "225",
    pages = "119",
    year = "1992"
}

@article{Benhar:1998au,
    author = "Benhar, Omar and Berti, Emanuele and Ferrari, Valeria",
    editor = "Ferrari, V. and Miller, J. C. and Rezzolla, L.",
    title = "{The Imprint of the equation of state on the axial w modes of oscillating neutron stars}",
    eprint = "gr-qc/9901037",
    archivePrefix = "arXiv",
    doi = "10.1046/j.1365-8711.1999.02983.x",
    journal = "Mon. Not. Roy. Astron. Soc.",
    volume = "310",
    pages = "797--803",
    year = "1999"
}

\end{document}